\documentclass[reprint,aps,prl,floatfix,nofootinbib,nobibnotes,
superscriptaddress]{revtex4-2}
\usepackage{amsmath,amssymb}
\usepackage{bm}
\usepackage{graphicx}
\usepackage{xcolor}
\usepackage[
colorlinks=true,
citecolor=blue,
urlcolor=blue,
linkcolor=red]{hyperref}

\begin{document}

\title{Interface-Driven Remote-Surface States in Rhombohedral Graphene}

\author{Pierre A. Pantale\'on}
\email{pierre.pantaleon@hainanu.edu.cn}
\affiliation{Center for Theoretical Physics, Hainan University, Haikou 570228, China}
\affiliation{School of Physics and Optoelectronic Engineering, Hainan University, Haikou 570228, China}

\author{Alejandro Jimeno-Pozo}
\altaffiliation[Present address: ]{Department of Physics and Astronomy, Uppsala University, Box 524, 751 20 Uppsala, Sweden}
\affiliation{Center for Theoretical Physics, Hainan University, Haikou 570228, China}
\affiliation{School of Physics and Optoelectronic Engineering, Hainan University, Haikou 570228, China}

\author{Francisco Guinea}
\email{paco.guinea@gmail.com}
\affiliation{Donostia International Physics Center, Paseo Manuel de Lardiz\'abal 4, San Sebasti\'an 20018, Spain}
\affiliation{IMDEA Nanoscience, C/ Faraday 9, 28049 Madrid, Spain}

\date{\today}

\begin{abstract}
Rhombohedral multilayer graphene hosts low-energy states on its two outer surfaces. We identify a layer-by-layer mechanism through which a low-energy state generated at one surface acquires weight on successive layers up to the opposite surface. For an outer-layer vacancy, the wavefunction spreads over a larger area on successive layers, shifting normalized weight toward the remote surface as the system size increases. The same relation governs the extension of sublattice-polarized zero modes bound to circular voids and straight zigzag terminations. At periodic moir\'e interfaces, it acts separately on each momentum harmonic, making the layer polarization of the low-energy bands depend on Bloch momentum and modulation period. This provides an intrinsic single-particle explanation for the remote-surface modulations observed in recent multilayer graphene experiments.
\end{abstract}

\maketitle

\textit{Introduction.---}Rhombohedral multilayer graphene hosts narrow electronic bands tunable by carrier density and electric displacement field. The low-energy states are concentrated on the two outer surfaces of the stack, and their dispersion reflects the chiral electronic structure of rhombohedral stacking~\cite{Min2008Chiral,Koshino2009Trigonal,Slizovskiy2019}. Alignment with hexagonal boron nitride (hBN) produces a moir\'e superlattice that reconstructs these surface bands and provides a platform for correlated insulating, superconducting, and topological phases~\cite{Chen2019,Chen2020,Park2023MultilayerHBN,Xie2025TunableFCI}. The substrate acts most directly on the adjacent graphene layer, whereas the low-energy states reside on both the adjacent and the opposite surface.

The influence of the moir\'e interface on the remote surface has been observed experimentally. NanoARPES of aligned rhombohedral pentalayer graphene reveals a moir\'e-enhanced flat band and replica bands~\cite{Zhang2025}. In rhombohedral trilayer graphene, scanning tunneling spectroscopy resolves spatially dependent features whose description requires an effective moir\'e potential on the surface away from the hBN interface~\cite{Seewald2025}. Scanning tunneling microscopy has also imaged trans-moir\'e orbitals on the remote surface of rhombohedral hexalayer graphene, which retain the periodicity of the distant interface and depend strongly on the graphene--hBN alignment~\cite{Wang2026TransMoire}. These observations show that an interface perturbation can leave a pronounced electronic signature across the stack.

Continuum calculations show that the spatial and layer character of moir\'e bands depends on displacement field, twist angle, and layer number~\cite{Park2023MultilayerHBN,Jimeno2025Topological}. Electron interactions have been proposed to transfer the interface modulation to the remote surface. In the moir\'e capacitor effect and the related charge-imprinting mechanism, a moir\'e-modulated charge distribution near the hBN interface generates an electrostatic potential for carriers on the opposite surface~\cite{Regnault2026MoireCapacitor,Qiao2026ChargeImprinting}, and interaction-driven charge redistribution has been invoked to describe trans-moir\'e orbitals~\cite{Wang2026TransMoire}. These interpretations place electron interactions at the center of the remote-surface response. However, the role of the single-particle connectivity of the rhombohedral stack in linking the interface to the opposite surface remains to be clarified.

The connection between states on different layers also appears in other configurations. A vacancy in monolayer graphene produces a quasilocalized state with an inverse-distance envelope~\cite{Pereira2006Disorder,Pereira2008Modeling}, while in Bernal bilayers and rhombohedral trilayers interlayer hopping adds components extending over neighboring layers~\cite{Castro2010Vacancy,Castro2012ABC}. Zigzag boundaries support sublattice-polarized states, whose structure depends on the stacking and termination~\cite{Castro2008_bilayer,Castro2008_graphite}, and analytic flat-band wavefunctions have been constructed for periodic chiral twisted graphene multilayers~\cite{Ledwith2022Family,Wang2022Hierarchy}. These results suggest a broader connection between the spatial structure of a state near one surface and its layer distribution, from isolated defects and boundaries to periodic interfaces.

In this Letter, we identify a microscopic layer-by-layer relation for low-energy states in rhombohedral multilayer graphene: a state generated in one surface acquires weight on successive layers up to the opposite surface. This relation follows from the chiral structure of the stack~\cite{Min2008Chiral,Koshino2009Trigonal}; here we show that it acts as a filter that continues an interface state, harmonic by harmonic, to the opposite surface. An outer-layer vacancy provides a real-space realization in which the charge weight shifts toward the remote layer as the system size changes. Circular voids and straight zigzag terminations reveal the roles of boundary geometry and penetration depth, respectively. For periodic moir\'e interfaces, the same relation acts on individual momentum harmonics, making the layer distribution depend on Bloch momentum and modulation period. Our results connect defect, boundary, and moir\'e states through the chiral connectivity of the stack and establish a general single-particle channel relevant to remote-surface moir\'e responses.

\begin{figure}[t]
\centering
\includegraphics[width=\columnwidth]{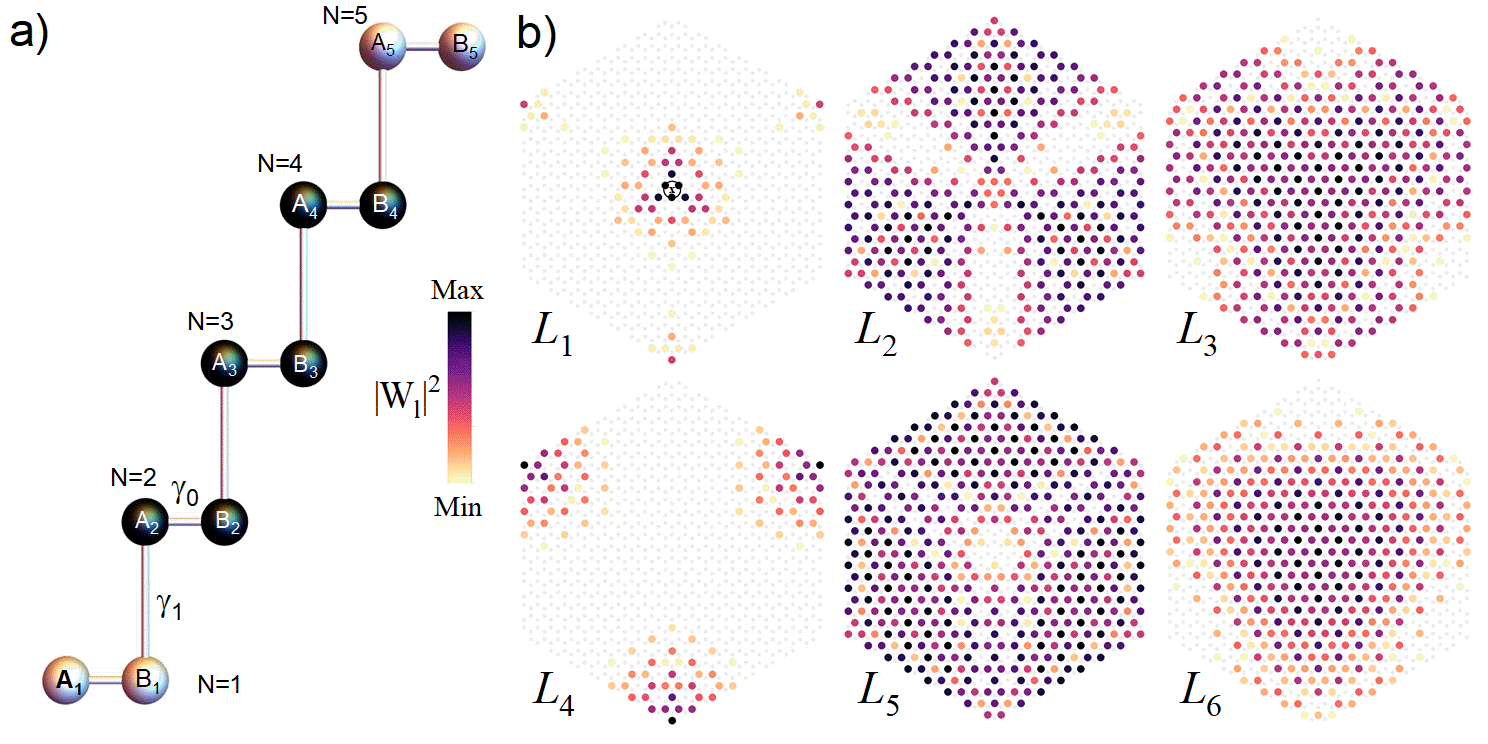}
\caption{\textbf{Lattice structure and vacancy-induced layer evolution.} (a) Minimal ABC chain for
$N=5$. The undimerized outer orbitals $A_1$ and $B_N$ are connected through
alternating intralayer terms and vertical hoppings $\gamma_1$. (b)
Layer-resolved lattice probability density of the zero mode produced by an
$A_1$ vacancy in an $N=6$ armchair flake. The vacancy is located at the origin
of $L_1$. Each layer is independently color-normalized to display its spatial
profile.}
\label{fig:MainTransfer}
\end{figure}

\textit{Rhombohedral stack and zero-mode propagation.---}
We consider an $N$-layer rhombohedral graphene stack with layers $L_1,\ldots,L_N$ and ABC stacking, with interlayer hopping $\gamma_1:B_l\leftrightarrow A_{l+1}$, where $l$ labels the layer. For a single valley, the low-energy multilayer Hamiltonian blocks are~\cite{Min2008Chiral,Koshino2009Trigonal}
\begin{equation}
\begin{aligned}
h(\bm p)&=\nu_0
\begin{pmatrix}
0&\pi^\dagger(\bm p)\\
\pi(\bm p)&0
\end{pmatrix},
&
T_{\gamma_1}&=
\begin{pmatrix}
0&0\\
\gamma_1&0
\end{pmatrix},
\end{aligned}
\label{eq:blocks-main}
\end{equation}
where $\pi(\bm p)=p_x+\mathrm{i}p_y$ and
$\nu_0=(\sqrt{3}/2)a\gamma_0$, with $a$ the graphene lattice constant and $\gamma_0$ the intralayer hopping. We set $\hbar=1$. The full Hamiltonian has diagonal blocks $H_{ll}=h$, nearest-layer blocks
$H_{l,l+1}=T_{\gamma_1}$, $H_{l+1,l}=T_{\gamma_1}^{\dagger}$, and vanishing remaining blocks. An illustration of the stack is shown in Fig.~\ref{fig:MainTransfer}(a).  

We consider a $B$-polarized zero mode generated in $L_1$, so that $\psi_{A_l}=0$ for every $l$, and define the first-layer component $\phi=\psi_{B_1}$. Because the Hamiltonian is bipartite, its zero-energy subspace can be resolved into sublattice-polarized components~\cite{Lieb1989Two}. The perturbation fixes the first-layer equation, which in an unmodified region of $L_1$ reduces to
\begin{equation}
\nu_0\pi^\dagger\phi=0.
\end{equation}
Irrespective of that first-layer equation, the $A_l$ rows for $l\geq2$ give
\begin{equation}
\nu_0\pi^\dagger\psi_{B_l}+\gamma_1\psi_{B_{l-1}}=0,
\qquad l=2,\ldots,N.
\label{eq:B-recursion-main}
\end{equation}
Combining the monolayer and interlayer terms gives a propagation rule
\begin{equation}
\widehat{\mathcal P}=-\frac{\gamma_1}{\nu_0}
(\pi^\dagger)^{-1},
\qquad
\psi_{B_l}=\widehat{\mathcal P}^{l-1}\phi.
\label{eq:propagator-operator-main}
\end{equation}
Equation~\eqref{eq:propagator-operator-main} isolates the component generated by $\phi$ and continues it from $L_1$ through the stack. For our stacking convention, neither an $A$-polarized state nor a state with $\psi_{B_1}=0$ enters this $B_1\rightarrow B_N$ channel.

\textit{Vacancy zero mode.---} Removing an $A_1$ site from a balanced stack creates a sublattice imbalance $N_B - N_A = 1$, where $N_A$ and $N_B$ denote the sublattice populations. This guarantees a zero-energy state on the $B$ sublattice~\cite{Lieb1989Two}, which can be chosen with $\psi_{A_l}=0$ in every layer, directly realizing Eqs.~\eqref{eq:B-recursion-main} and~\eqref{eq:propagator-operator-main}. At one Dirac point, using $z=x+\mathrm{i}y$, $\bar z=x-\mathrm{i}y$, and $\pi^\dagger=-2\mathrm{i}\partial_z$, the monolayer vacancy solution is of the form $\psi_{B_1}(z,\bar z)=C/\bar z$~\cite{Pereira2006Disorder}. Successive integration of Eq.~\eqref{eq:B-recursion-main} gives the solution (see Sec.~S3 of the Supplemental Material):
\begin{equation}
\psi_{B_l}(z,\bar z)=
\frac{C(-\mathrm{i})^{l-1}}{(l-1)!}
\frac{z^{l-1}}{\ell_\perp^{l-1}\bar z},
\qquad
\ell_\perp=\frac{2\nu_0}{\gamma_1}.
\label{eq:vacancy-solution-main}
\end{equation}
Writing $z^{l-1}/\bar z=r^{l-2}\exp(\mathrm{i}l\varphi)$, where $\varphi$ is the polar angle, reveals a real-space phase winding of $l$. This real-space winding is the real space analogous of  the momentum-space pseudospin winding that enters the multilayer Berry phase~\cite{Park2011Berry}. The corresponding envelope is
\begin{equation}
|\psi_{B_l}(r)|=
\frac{|C|}{r}\frac{(r/\ell_\perp)^{l-1}}{(l-1)!}.
\label{eq:vacancy-envelope-main}
\end{equation}
Although valley and basis conventions alter the phase, the envelope remains invariant: the first layer retains the monolayer $1/r$ form, the second-layer magnitude is independent of $r$, and the $l$-th layer magnitude scales as $r^{l-2}$. For a Bernal bilayer ($N=2$), this reproduces the known extended component in Ref.~\cite{Castro2010Vacancy}; for a rhombohedral trilayer ($N=3$), the $z^2/\bar z$ dependence is consistent with the extended outer-layer state in Ref.~\cite{Castro2012ABC}.

The single-valley amplitude gains one unit of phase winding per layer and belongs to a $C_3$ representation determined by $l\bmod 3$. Layers separated by three steps share this spatial symmetry, though not their radial dependence or pointwise charge density. This symmetry classification is consistent with the recurrence in Fig.~\ref{fig:MainTransfer}(b) (see Sec.~S3 of the Supplemental Material).

\textit{Finite-size effects.---}%
Because the higher-layer components of Eq.~\eqref{eq:vacancy-envelope-main} grow radially, the normalized weights depend on the system size.\ We illustrate this effect using finite armchair multilayer flakes, as shown in Fig.~\ref{fig:MainTransfer}(b) for $N=6$, whose boundary suppresses the zero-energy zigzag modes~\cite{Wakabayashi2010Electronic} while retaining the rhombohedral stack configuration~\cite{Heikkil2011Dimensional,Heikkil2011Flat}. This isolates the vacancy state from zigzag boundary modes. 

Let $R$ denote the radial scale defined by the lattice construction (see Sec.~S3 of the Supplemental Material), and define
\begin{equation}
X=\frac{R}{\ell_\perp}
=\frac{R\gamma_1}{3a_{cc}\gamma_0},
\qquad a=\sqrt{3}a_{cc},
\label{eq:radius-main}
\end{equation}
where $a_{cc}$ is the carbon--carbon distance. At a common radius, Eq.~\eqref{eq:vacancy-envelope-main} yields
\begin{equation}
\frac{|\psi_{B_{l+1}}(r)|}{|\psi_{B_l}(r)|}
=\frac{r}{l\ell_\perp},
\qquad
\frac{|\psi_{B_N}(R)|^2}{|\psi_{B_1}(R)|^2}
=\left[\frac{X^{N-1}}{(N-1)!}\right]^2.
\label{eq:layer-ratios-main}
\end{equation}
For this particular solution, the amplitude at $r=R$ decreases at each layer step when $X<1$; for $X>1$, it initially increases and continues to do so from $L_l$ to $L_{l+1}$ while $X>l$. The integrated weights for $R\gg r_0$ are 
\begin{equation}
\begin{aligned}
w_1&\simeq |C|^2\ln(R/r_0),\\
w_l&\simeq
\frac{|C|^2X^{2l-2}}{(2l-2)[(l-1)!]^2},
\qquad l\geq2,
\end{aligned}
\label{eq:weights-main}
\end{equation}
The infinite-plane state is intrinsically non-normalizable: $w_1$ diverges logarithmically~\cite{Pereira2006Disorder}, while every subsequent $w_{l\geq2}$ diverges as a power of $R$. Consequently, for fixed $N$, increasing $R$ transfers weight toward the remote layer $L_N$ while redistributing the zero-mode density across the flake. 

Finite-lattice calculations confirm this transfer. The normalized probability $W_l$ in layer $L_l$ (see Sec.~S3 of the Supplemental Material) shifts steadily from $L_1$ toward $L_N$ as $R$ grows. For $N=6$ (Table~S1 of the Supplemental Material), at $X=1.41$ the remote layer already outweighs the vacancy layer ($W_6=25.3\%$ versus $W_1=17.3\%$), although an intermediate layer carries the maximum. At $X=1.75$, $W_6=48.5\%$ while $W_1=4.1\%$. A perturbation in $L_1$ thus reaches $L_N$ with substantial weight, which dominates for larger flakes.

The pristine flakes exhibit the usual low-energy surface-state pair~\cite{Heikkil2011Flat, Heikkil2011Dimensional}.  Boundary hybridization splits these states to nonzero energies; the splitting collapses as $R$ increases (Sec.~S3 of the Supplemental Material). An $A_1$ vacancy instead pins an exact zero mode for all sizes. As $R$ grows, its weight shifts toward $L_N$, acquiring remote-$B_N$ character as the pristine pair approaches zero energy. Figure~\ref{fig:MainTransfer}(b) shows the charge distribution for a vacancy at the origin of $L_1$, with independent layer color normalization highlighting each spatial profile.

\textit{Circular void.---} Our previous analysis extends to an ideal circular void of radius $R_0$ in $L_1$ with zigzag boundary conditions. In the same single-valley convention, the first-layer solution is (see Sec.~S4 of the Supplemental Material)
\begin{equation}
\psi_{B_1}^{(m)}(z,\bar z)=
\begin{cases}
0, & r\leq R_0,\\
C_m/\bar z^m, & r>R_0,
\end{cases}
\qquad m=1,2,\ldots.
\end{equation}
Recursive continuation with Eq.~\eqref{eq:propagator-operator-main} gives the component that vanishes beneath the void in subsequent layers,
\begin{equation}
\psi_{B_l}^{(m)}(z,\bar z)=
\begin{cases}
0, & r\leq R_0,\\[2pt]
\displaystyle
\frac{C_m(-\mathrm{i})^{l-1}}{(l-1)!\ell_\perp^{l-1}}
\frac{\left(z-R_0^2/\bar z\right)^{l-1}}{\bar z^m},
& r>R_0.
\end{cases}
\end{equation}
Only $L_1$ contains a physical hole. Nevertheless, the propagated component retains a density depletion beneath the void, producing a
wavefunction {\it scar} in subsequent layers up to the remote surface. This {\it remote imprint} follows from the factor $z-R_0^2/\bar z$, which
vanishes at the void boundary. Far from the void, $|\psi_{B_l}^{(m)}|\propto r^{l-1-m}$, so each layer step adds one
radial power and channels with smaller $m$ extend farther. The angular index $m$, the scale $R_0/\ell_\perp$, and any finite outer boundary
therefore determine the layer distribution. Conformal mappings can construct first-layer seeds for other shapes, while their multilayer
continuation must follow from the layer equations in physical coordinates~\cite{katsnelson_transport_2008,rycerz_conformal_2009}.
For derivations and lattice descriptions see Sec.~S4 of the Supplemental Material.

\textit{Zigzag zero modes.---} We now consider a zigzag step parallel to $x$: $L_1$ occupies $y>0$, while the other layers span the full plane. This termination also supports sublattice-polarized modes~\cite{Fujita1996Peculiar,Castro2008_bilayer,
Castro2008_graphite}. An $A_1$ edge occupies the undimerized $A_1$ orbital, has no vertical neighbor, and remains localized in $L_1$ within the minimal model. A $B_1$ edge, however, generates a $B_1$-polarized mode in the $B_1\rightarrow B_N$ channel directly realizing Eqs.~\eqref{eq:B-recursion-main} and~\eqref{eq:propagator-operator-main}. In the continuum, the decaying $B_1$ solution has an inverse penetration length $\kappa>0$ and a longitudinal momentum $-\kappa$, given by
\begin{equation}
\psi_{B_1}(x,y;\kappa)=
\left\{
\begin{array}{lr}
0, & y\leq0,\\
C_\kappa \mathrm e^{-\mathrm i\kappa x-\kappa y}, & y>0.
\end{array}
\right.
\label{eq:zigzag-seed-main}
\end{equation}
Applying Eq.~\eqref{eq:B-recursion-main} at fixed $\kappa$ yields
\begin{equation}
\psi_{B_l}(x,y;\kappa)=
\left\{
\begin{array}{lr}
0, & y\leq0,\\
\displaystyle
\frac{C_\kappa}{(l-1)!}
\left(\frac{\gamma_1y}{\nu_0}\right)^{l-1}
\mathrm e^{-\mathrm i\kappa x-\kappa y}, & y>0,
\end{array}
\right.
\label{eq:zigzag-solution-main}
\end{equation}
for $l=1,\ldots,N$. The components vanish for $y<0$ and smoothly match zero at $y=0$ for $l\geq2$; only $L_1$ has a physical edge. The exponential factor fixes the penetration depth, and each layer step adds a power of $y$, producing a continuous one-dimensional zero-energy mode. 

On the lattice, the edge momentum $q$ has a decay factor $\lambda_q=-2\cos(q/2)$, with normalizability requiring $|\lambda_q|<1$, or $2\pi/3<q<4\pi/3$ modulo $2\pi$~\cite{Fujita1996Peculiar} (see Sec.~S5 of the Supplemental Material). Thus, $q$ labels the mode, and $|\lambda_q|$ determines its penetration. With spacing $a_\perp=3a_{cc}/2$ between consecutive zigzag rows 
perpendicular to the edge, the inverse decay length is
$\kappa(q)=-a_\perp^{-1}\ln|\lambda_q|$. Hence, $\kappa(q)\rightarrow0$ as $|\lambda_q|\rightarrow1$ at the interval boundaries, whereas $\kappa(q)\rightarrow\infty$ at $q=\pi$, where $\lambda_q=0$. The integrated norm per unit edge length follows from Eq.~\eqref{eq:zigzag-solution-main}:
\begin{equation}
\frac{w_l(\kappa)}{w_1(\kappa)}=
\binom{2l-2}{l-1}
\left(\frac{1}{\kappa\ell_\perp}\right)^{2l-2},
\qquad
\ell_\perp=\frac{2\nu_0}{\gamma_1}.
\label{eq:zigzag-weight-main}
\end{equation}
Thus, the same layer-by-layer propagation connects a zigzag boundary state to the remote surface, with the penetration depth $\kappa^{-1}$ controlling its distribution across the stack. Modes penetrating farther into the first layer acquire a larger fraction of their weight on the remote surface.

\begin{figure}[t]
\centering
\includegraphics[width=\columnwidth]{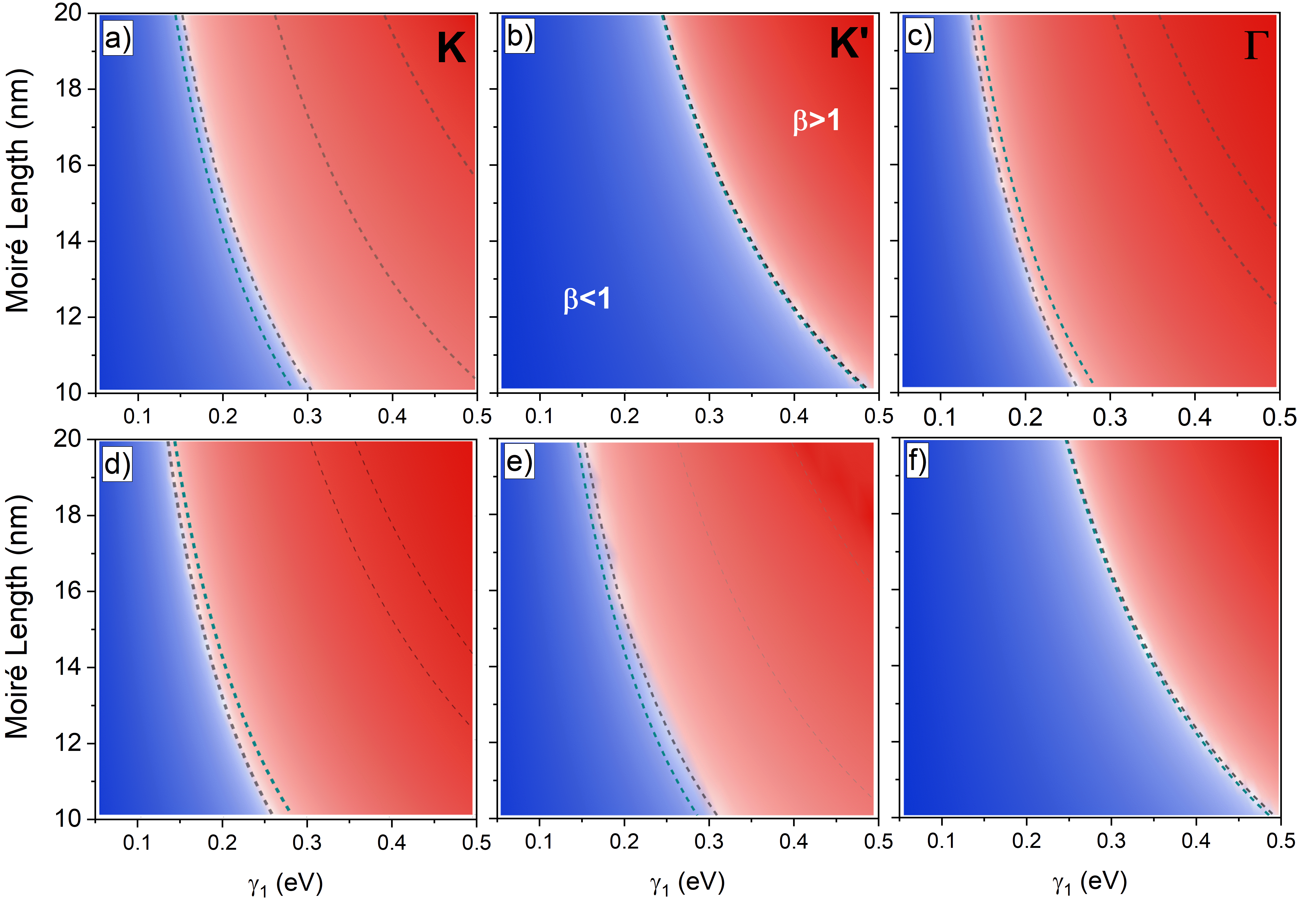}
\caption{\textbf{Momentum-resolved layer crossover.} Maps of $\beta$ versus
$\gamma_1$ and the superlattice period for the pseudogauge system [(a)--(c)]
and the twisted-monolayer system [(d)--(f)]. Columns correspond to $K$, $K'$,
and $\Gamma$. Blue and red denote $\beta<1$ and $\beta>1$, respectively. The
white region and black dashed curve mark $\beta=1$ for
$\bm G_{-1,-1}=-(\bm G_1+\bm G_2)$; gray dashed curves show other harmonic channels. The contours follow Eq.~\eqref{eq:boundary-main}.}
\label{fig:Btransition}
\end{figure}

\textit{Momentum-resolved layer propagation.---} The vacancy and boundary states illustrate how the spatial extent of
a perturbation in $L_1$ influences the weight that reaches the remote
surface. We now extend this analysis to periodic moir\'e systems. Here, an interface state contains several momentum harmonics, each continued through the rhombohedral stack according to the
same layer relation.

For a nonzero momentum $\bm p$, the low-energy surface projection
relates the interface component $B_1$ to the remote surface $B_N$
(see Sec.~S6 of the Supplemental Material),
\begin{eqnarray}
\psi_{B_1}(\bm p)&=&D_{N-1}^{\dagger}(\bm p)
\psi_{B_N}(\bm p),\\
D_{N-1}(\bm p)&=&
\left[-\frac{\nu_0\pi(\bm p)}{\gamma_1}\right]^{N-1}.
\label{eq:D-main}
\end{eqnarray}
This relation is the momentum-space counterpart of the layer recursion in Eq.~\eqref{eq:B-recursion-main}. We define
the layer-step magnitude
\begin{equation}
\beta(\bm p)=
|D_{N-1}(\bm p)|^{-1/(N-1)}
=\frac{\gamma_1}{\nu_0|\bm p|},
\label{eq:beta-main}
\end{equation}
which gives
\begin{equation}
\frac{|\psi_{B_N}(\bm p)|}{|\psi_{B_1}(\bm p)|}
=\beta^{N-1}(\bm p).
\label{eq:momentum-ratio-main}
\end{equation}
Equation~\eqref{eq:momentum-ratio-main} establishes one of our central results: for low-energy states, the ABC stack acts as a momentum-selective channel connecting the interface to the remote surface. For each nonzero momentum, the amplitude in the $B_1\rightarrow B_N$ channel decreases across the stack when
$\beta(\bm p)<1$ and increases when $\beta(\bm p)>1$. Since $\beta(\bm p)\propto|\bm p|^{-1}$, long-wavelength harmonics are preferentially transferred: their remote-to-interface amplitude ratio scales as $|\bm p|^{-(N-1)}$. The exponent $N-1$ thus makes the momentum selectivity more pronounced in thicker stacks. The condition $\beta(\bm p)=1$ marks the crossover for an individual harmonic, while the overall layer polarization depends on the combined weights of all harmonics populated by the interface. 

\begin{figure*}
\centering
\includegraphics[width=0.92\textwidth]{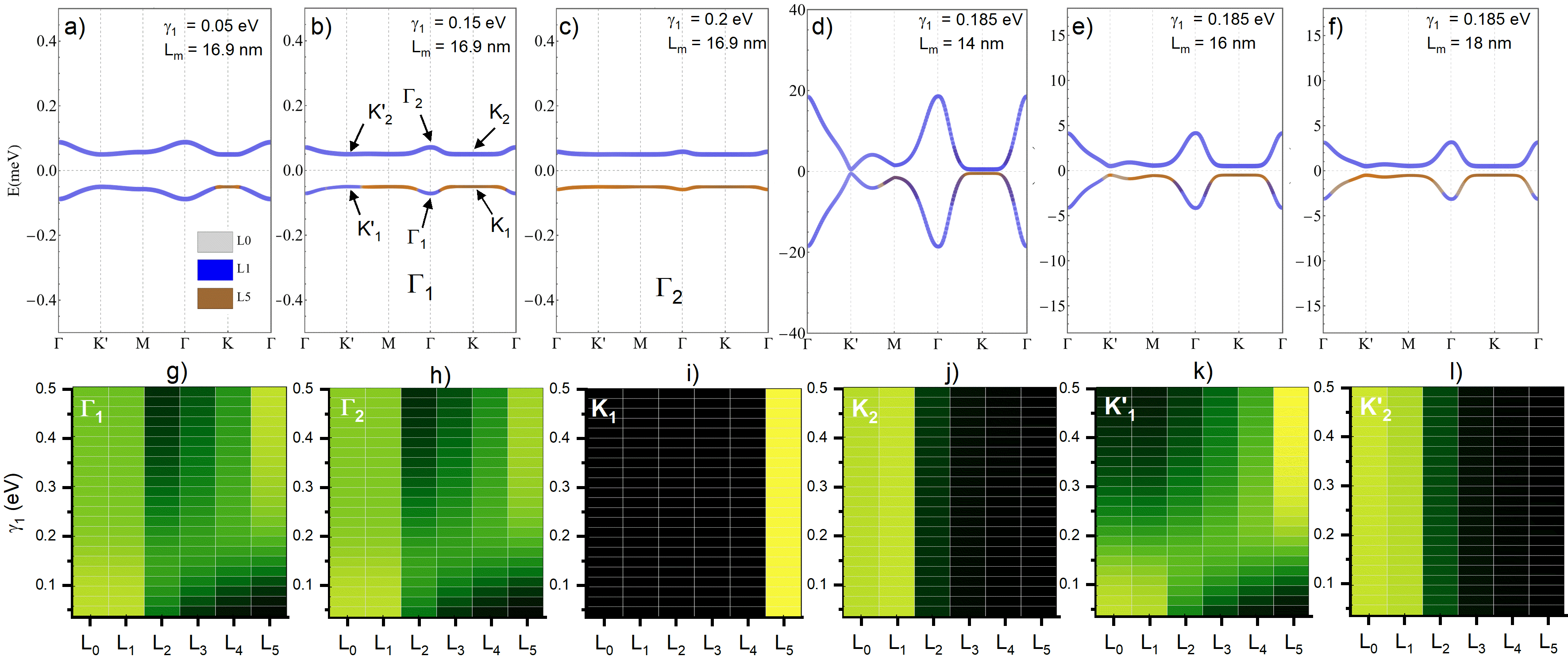}
\caption{\textbf{Layer redistribution in twisted
monolayer--rhombohedral graphene.} Band structures for an $N+1$
system with $N=5$. A small uniform sublattice mass $m_0\sigma_z$ is included on every layer. The calculation uses $|m_0|=0.05$ meV, $\gamma_0=3.16$ eV, and $u'=0.0975$ eV. Panels (a)--(c) use $L_M=16.9$ nm and $\gamma_1=0.05$, $0.15$, and $0.20$ eV; panels (d)--(f) use $\gamma_1=0.185$ eV and $L_M=14$, $16$,
and $18$ nm. Colors resolve the weights from the interface layers $L_0,L_1$ to the remote surface $L_5$. At $L_M=16.9$ nm, panels
(g)--(l) show the layer weights of the lower (lower index 1) and upper (lower index 2) central states at $\Gamma$, $K$, and $K'$ versus $\gamma_1$. The lower state evolves
toward $L_5$, whereas the upper state remains predominantly interface-polarized.}
\label{fig:BandasTwisted}
\end{figure*}

To illustrate this momentum-selective propagation, we consider two periodic interface systems. The first is a transverse pseudogauge field acting on $L_1$,
\begin{equation}
\begin{aligned}
V_{\mathrm g}(\bm r)&=
\left[\widehat{\bm z}\times
\frac{u_\perp\nabla f(\bm r)}{G}\right]\cdot\bm\sigma,\\
f(\bm r)&=-2\sum_{j=1}^{3}\sin(\bm g_j\cdot\bm r),
\end{aligned}
\label{eq:gauge-main}
\end{equation}
where $|\bm g_j|=G=4\pi/(\sqrt{3}L_{\mathrm M})$ and $L_{\mathrm M}$
is the moiré length. Spatially varying strain can generate
pseudogauge fields in graphene~\cite{Vozmediano2010Gauge,
Guinea2010Strain}, and periodic fields can produce narrow central
bands~\cite{Gao2023Strain,Wan2023Nearly,SanJose2014Electronic,
PabloSanJose2014Spontaneous}. The smooth, zero-flux field considered
here supports zero-energy states at the folded Dirac point in the
minimal chiral model. Away from this point, the central bands are
generally dispersive.

The second realization is a monolayer $L_0$ twisted relative to the
rhombohedral stack $L_1,\ldots,L_N$~\cite{Jimeno2025Topological}.
In the chiral limit, the tunneling between $L_0$ and $L_1$ is
\begin{equation}
T_{01}(\bm r)=\sum_{j=1}^{3}T_j
\mathrm{e}^{-\mathrm{i}\bm Q_j\cdot\bm r},
\qquad
T_j=u'
\begin{pmatrix}
0&\mathrm{e}^{-\mathrm{i}\phi_j}\\
\mathrm{e}^{\mathrm{i}\phi_j}&0
\end{pmatrix},
\label{eq:twist-main}
\end{equation}
where $\phi_j=2\pi(j-1)/3$ and $u'$ is the interlayer tunneling amplitude. At the chiral magic angles~\cite{Tarnopolsky2019_origin,
Khalaff2019Magic,Grisha2023Magic,Naumis2021Reduction}, this system inherits two exact zero-energy flat bands. The magic angles are
independent of $N$ and $\gamma_1$~\cite{Ledwith2022Family,Wang2022Hierarchy}. One state remains in the interface-connected $A$ sector, while the other has a $B_1$ component continued through the rhombohedral stack.

In both systems, the periodic interface couples different harmonics. For a Bloch momentum $\bm k$, the propagation momenta in the periodic pseudogauge and twisted moiré systems are, respectively,  
\begin{equation}
\bm p_i^{\mathrm g}=\bm k+\bm G_i,
\qquad
\bm p_i^{\mathrm{tw}}=\bm k-\bm K_R+\bm G_i,
\label{eq:momenta-main}
\end{equation}
where $\bm G_i$ are reciprocal superlattice vectors and $\bm K_R$ is the folded Dirac point of the rhombohedral slab in the twisted system. Each harmonic therefore has its own $\beta(\bm p_i)$. The interface determines the harmonic content of the state, while the ABC connectivity determines the continuation of its zero-energy $B_1$ component through the stack.

The modulation period sets the reciprocal momenta and hence the propagation scale. Consider a triangular moir\'e interface with
period $L_M$ and primitive reciprocal vectors $\bm G_1,\bm G_2$ of length $4\pi/(\sqrt{3}L_M)$. For a
Bloch momentum $\bm k$, we write $\bm k-\bm K_R=c_1\bm G_1+c_2\bm G_2$ and $\bm G_{mn}=m\bm G_1+n\bm G_2$. Here $\bm K_R$ is the folded
Dirac point of the ABC stack, $c_{1,2}$ are dimensionless momentum coordinates, and $m,n\in\mathbb Z$ label the
harmonics. For example, $\bm K_R=\bm K$ for the twisted interface.

For fixed $c_{1,2}$, the condition $\beta(\bm k-\bm K_R+\bm G_{mn})=1$ gives the crossover period
\begin{equation}
\begin{aligned}
L_M^*&=\frac{4\pi\nu_0}{\sqrt{3}\gamma_1}
\sqrt{\mathcal Q_{mn}},\\
\mathcal Q_{mn}&=(m+c_1)^2+(n+c_2)^2
-(m+c_1)(n+c_2),
\end{aligned}
\label{eq:boundary-main}
\end{equation}
for a nonzero-momentum harmonic, $\mathcal Q_{mn}>0$. Note that this result is independent of the microscopic interface coupling: the interface selects and weights the harmonics, while the ABC stack determines their propagation. For fixed reduced momentum, the crossover of an individual harmonic is independent of $N$, whereas its remote-to-interface amplitude ratio scales as $\beta^{N-1}$. For the first-star harmonics at $\bm k=\bm K_R$ with $\mathcal Q_{mn}=1$, Eq.~\eqref{eq:boundary-main} gives $L_M^*=4\pi\nu_0/(\sqrt3\gamma_1)\approx13.6$~nm for $\gamma_1=0.36$~eV ($\approx12.9$~nm for $\gamma_1=0.38$~eV). This is close to the maximal period of aligned graphene/hBN, $\approx14$~nm. Aligned devices therefore lie near the crossover for the first-star harmonics, while all higher harmonics have $\beta<1$. This filtering yields three signatures. (i) Harmonics are suppressed on the remote surface relative to the interface by $\beta^{N-1}$, so the second-star harmonics ($|\bm G|=\sqrt3G$) are reduced relative to the first star by an additional factor $3^{-(N-1)/2}$ ($\approx0.06$ for $N=6$): the remote-surface texture is a low-pass-filtered, smoother copy of the interface modulation. (ii) This suppression strengthens with layer number. (iii) Since misalignment shortens $L_M$, the first-star harmonics cross $\beta=1$ near $L_M^*$, and the remote modulation should drop sharply with graphene--hBN misalignment, more steeply for thicker stacks; this is consistent with the alignment sensitivity reported in Ref.~\cite{Wang2026TransMoire}. This effect is summarized in Fig.~\ref{fig:Btransition} where by increasing $\gamma_1$ or the modulation period increases $\beta$ for a given nonzero harmonic, enhancing its relative amplitude on the remote surface. Different harmonics and Bloch momenta reach $\beta=1$ at different parameters. This momentum selectivity allows an interface modulation to produce a spatial imprint on the opposite surface. In particular, the moir\'e potential generated by an hBN substrate contains scalar, sublattice-mass, and pseudogauge components~\cite{SanJose2014Electronic,PabloSanJose2014Spontaneous,Park2023MultilayerHBN}. These components shape the interface states and their harmonic content, while the ABC connectivity allows the interface-induced wavefunctions to extend through the stack. This
mechanism is therefore relevant to recent trans-moir\'e features observed on the remote surface~\cite{Wang2026TransMoire}. The condition $\beta=1$ applies to an individual harmonic; the total layer distribution depends on the complete state and on the coupling
between its harmonics. The scale $\beta(\bm p)$ is defined for nonzero momentum and follows from the zero-energy layer relation. For the dispersive bands, we calculate the layer weights from the full Hamiltonian and use $\beta$ to interpret their momentum-dependent evolution (see Sec.~S6 of the Supplemental Material). 

The twisted-system bands in Fig.~\ref{fig:BandasTwisted} show that increasing $\gamma_1$ or $L_M$ can move the lower central state from $L_0,L_1$ toward $L_5$, with appreciable intermediate-layer weight and momentum-dependent distribution. The upper central state remains predominantly interface-polarized in the displayed spectra.  The pseudogauge system shows a related momentum dependence: the central states at $\Gamma$ remain polarized on opposite surfaces, while the lower state at $K$ evolves from $L_1$ toward $L_N$ as $\gamma_1$ increases (see Sec.~S6 of the Supplemental Material).

At each moiré length, the full Hamiltonian is rediagonalized. The resulting layer weights therefore include both the change in harmonic content of the interface state and its continuation through the stack. Their coherent superposition also determines the remote-surface textures, which can be strongly modulated or nearly uniform depending on momentum (see Sec.~S6 of the Supplemental Material). By reshaping the remote-surface wavefunctions, momentum-selective propagation modifies the projected Coulomb interaction, providing a microscopic link between the moir\'e interface and the interaction-driven fractional Chern phases~\cite{Sarkar2026Similarity}. We note that the intrinsic single-particle mechanism identified here provides a direct route for interface-induced modulation of the remote surface. Electronic interactions act on the resulting states and can generate an additional remote-surface potential through the proposed moir\'e capacitor and charge-imprinting mechanisms~\cite{Regnault2026MoireCapacitor,Qiao2026ChargeImprinting}, thereby reshaping the response.

\textit{Trigonal warping.---} The analysis presented so far does not take into account the effect of interlayer hoppings that break the isotropy of the model and induce trigonal warping, usually denoted by $\gamma_3$ and $\gamma_4$. These hoppings lead to momentum-dependent terms in the Hamiltonian and vanish as $|\bm k|\rightarrow0$. As mentioned above, the perturbations projected from one surface to the other have low momenta, so we expect these interlayer terms to have a weak effect on the phenomena discussed previously. We obtain a more quantitative analysis by projecting these hoppings onto the two low-energy wavefunctions (Supplemental Material, Sec.~S7). The $\gamma_3$ term describes hoppings between neighboring layers and different sublattices, so it hybridizes the two surface-localized states. The term proportional to $\gamma_4$ does not hybridize the sublattices. We obtain
\begin{equation}
\begin{aligned}
t_3(\bm k)&=\frac{A_{\bm k}}{|A_{\bm k}|}\mathcal C_{\bm k}^{-2}
\nu_3 k\mathrm e^{\mathrm i\varphi_{\bm k}}\rho_{\bm k}^{N-2}\\
&\quad\times\frac{1-\mathrm e^{2\mathrm i(N-1)\varphi_{\bm k}}}
{1-\mathrm e^{2\mathrm i\varphi_{\bm k}}},\\
\varepsilon_4(\bm k)&=\mathcal C_{\bm k}^{-2}\nu_4 k
\mathrm e^{\mathrm i\varphi_{\bm k}}\rho_{\bm k}
\frac{1-|\rho_{\bm k}|^{2(N-1)}}{1-|\rho_{\bm k}|^2}.
\end{aligned}
\label{eq:warping-main}
\end{equation}
Here $k=|\bm k|$, $\varphi_{\bm k}=\arg(k_x+\mathrm i k_y)$, $\nu_j=(\sqrt{3}/2)a\gamma_j$ for $j=3,4$, $\rho_{\bm k}=-\gamma_1/(\nu_0 k\mathrm e^{\mathrm i\varphi_{\bm k}})$, and $\mathcal C_{\bm k}^2=(1-|\rho_{\bm k}|^{2N})/(1-|\rho_{\bm k}|^2)$. The first-layer amplitude $A_{\bm k}$ and matrix elements are defined in Sec.~S7. The first-order perturbative correction considered here changes the energies of the eigenstates, but not their wavefunctions (which are hybridized by the $\gamma_3$ term), so that the relation between amplitudes on the top and bottom surfaces remains unaltered.

{\it Conclusions.\textendash} We establish an intrinsic interface-to-surface propagation mechanism in multilayer graphene. Once a perturbation generates or couples to the appropriate sublattice-polarized interface component, the chiral ABC equations continue its wavefunction through the stack. A surface-confined perturbation can thus produce wavefunction weight and charge density on the opposite surface without acting there directly.

Vacancies, circular voids, and straight zigzag boundaries reveal complementary real-space controls: lateral system size governs vacancy-state redistribution, discrete angular channels determine how the scar generated by the void extends through the stack, and penetration depth controls the straight boundary mode distribution. At periodic interfaces, the propagation is momentum dependent, with lower-momentum components reaching remote layers more effectively.  The modulation period, Bloch momentum, and interlayer coupling determine how each harmonic is redistributed across the stack.

These results microscopically connect the two surfaces of substrate-coupled and twisted rhombohedral systems~\cite{Park2023MultilayerHBN,Zhang2025,Seewald2025, Wang2026TransMoire}. Substrate potentials, lattice relaxation, electrostatic reconstruction, and electronic correlations can modify the observable remote-surface electronic structure~\cite{Regnault2026MoireCapacitor,Qiao2026ChargeImprinting}. Defects, boundaries, and moir\'e interfaces thus realize one principle: rhombohedral connectivity converts the spatial and momentum scales of a surface perturbation into layer polarization, providing control over wavefunction weight and charge across the multilayer.

{\it Acknowledgments.\textendash} We thank Danna Liu and Arnob Kumar Ghosh for discussions. This research is supported by the National Natural Science Foundation of China (NSFC) grant No.\ 12547103.\ IMDEA Nanociencia acknowledges support from the ‘Severo Ochoa’ Programme for Centres of Excellence in R\&D (CEX2020-001039-S/AEI/10.13039/501100011033). P.A.P. and F.G. acknowledge support from NOVMOMAT, project PID2022-142162NB-I00 funded by MICIU/AEI/10.13039/501100011033 and by FEDER, UE as well as financial support through the (MAD2D-CM)-MRR MATERIALES AVANZADOS-IMDEA-NC. F.G. acknowledges funding from the EU NextGenerationEU/PRTR-C17.I1, as well as by the IKUR Strategy under the collaboration agreement between Ikerbasque Foundation and DIPC on behalf of the Department of Education of the Basque Government. A.J.-P. acknowledges the Center for Theoretical Physics and the School of Physics and Optoelectronic Engineering at Hainan University for their hospitality, where this work was developed during his visit.

\bibliography{references}

\onecolumngrid
\newpage

\setcounter{equation}{0}
\setcounter{figure}{0}
\setcounter{table}{0}
\setcounter{page}{1}
\setcounter{section}{0}

\renewcommand{\thesection}{S\arabic{section}}
\renewcommand{\theequation}{S\arabic{equation}}
\renewcommand{\thetable}{S\arabic{table}}
\renewcommand{\thefigure}{S\arabic{figure}}
\setcounter{secnumdepth}{3}
\renewcommand{\thesubsection}{\Alph{subsection}}
\renewcommand{\thesubsubsection}{\thesubsection.\arabic{subsubsection}}

\begin{center}
\Large Supplementary Information for \\
``Interface-Driven Remote-Surface States in Rhombohedral Graphene''
\end{center}
\begin{center}
\normalsize Pierre A. Pantale\'on, Alejandro Jimeno-Pozo and Francisco Guinea
\end{center}

\section{Geometry and minimal ABC Hamiltonian}
\label{sec:geometry}

The rhombohedral slab contains layers $L_1,\ldots,L_N$. Each graphene layer is
written in the basis $(A_l,B_l)$. Our ABC convention is given by $\gamma_1:B_l\longleftrightarrow A_{l+1}$ with $l=1,\ldots,N-1$, so that $A_1$ and $B_N$ are the undimerized sites. For a single valley, we have
\begin{equation}
h(\bm p)=\nu_0
\begin{pmatrix}
0&\pi^\dagger(\bm p)\\
\pi(\bm p)&0
\end{pmatrix},
\qquad
T_{\gamma_1}=
\begin{pmatrix}
0&0\\
\gamma_1&0
\end{pmatrix}\textcolor{red}{,}
\label{eq:blocks}
\end{equation}
where $\pi(\bm p)=p_x+\mathrm{i}p_y$ with $\nu_0=\frac{\sqrt{3}}{2}a\gamma_0$. Here $a$ is the graphene lattice constant, $\gamma_0$ is the intralayer hopping, and $\hbar=1$. 
The multilayer Hamiltonian is
\begin{equation}
H_N^{\mathrm{ABC}}(\bm p)=
\begin{pmatrix}
h&T_{\gamma_1}&0&\cdots&0\\
T_{\gamma_1}^\dagger&h&T_{\gamma_1}&\ddots&\vdots\\
0&T_{\gamma_1}^\dagger&h&\ddots&0\\
\vdots&\ddots&\ddots&\ddots&T_{\gamma_1}\\
0&\cdots&0&T_{\gamma_1}^\dagger&h
\end{pmatrix}.
\label{eq:HABC}
\end{equation}
This is the minimal continuum form of rhombohedral graphene
\cite{Min2008Chiral,Koshino2009Trigonal}. 

\section{General zero-mode propagation through the ABC stack}
\label{sec:general-recursion}

At zero energy, chiral symmetry allows the wavefunction to be resolved into
sublattice-polarized components. Consider a $B$-polarized component generated
near $L_1$, so that $\psi_{A_l}=0$ and
$\phi=\psi_{B_1}$. In real space, introduce $z=x+\mathrm{i}y$ and
$\bar z=x-\mathrm{i}y$. With $p_j\rightarrow-\mathrm{i}\partial_j$,
\begin{equation}
\pi^\dagger=-2\mathrm{i}\partial_z.
\label{eq:real-pi}
\end{equation}
The perturbation determines the first-layer equation obeyed by $\phi$. Where $L_1$ is unperturbed it reduces to $\nu_0\pi^\dagger\phi=0$. Independently of that interface equation, the $A_l$ rows for $l\geq2$ are
\begin{equation}
\nu_0\pi^\dagger\psi_{B_l}
+\gamma_1\psi_{B_{l-1}}=0,
\qquad l=2,\ldots,N,
\label{eq:continuum-recursion}
\end{equation}
which is the layer recursion quoted in the main text. It defines the
$B_1\rightarrow B_N$ propagation channel. It is
useful to write the layer step in the compact form
\begin{equation}
\widehat{\mathcal P}
=-\frac{\gamma_1}{\nu_0}
(\pi^\dagger)^{-1},
\qquad
\psi_{B_l}=
\widehat{\mathcal P}^{l-1}\psi_{B_1}.
\label{eq:P-operator}
\end{equation}
The interface determines $\phi$, while the ABC Hamiltonian determines the local continuation. An exact zero mode in the $B_1$-connected chiral sector therefore satisfies
\begin{equation}
\psi_{B_l}^{[\phi]}=\widehat{\mathcal P}^{l-1}\phi,
\qquad l=1,\ldots,N,
\label{eq:arbitrary-seed}
\end{equation}
provided that no independent zero-energy component begins in a later layer. This conditional propagation law is the general result used in the concrete realizations below.

\section{Vacancy-induced propagation}
\label{sec:vacancy}
\subsection{Lattice zero mode}
\label{sec:vacancy-lattice}

Removing an $A_1$ site from a sublattice-balanced pristine stack gives the global sublattice imbalance $N_B-N_A=1$.  The bipartite nearest-neighbor Hamiltonian therefore has at least one zero mode on the majority $B$ sublattice
\cite{Pereira2006Disorder,Pereira2008Modeling}.  We select a
vacancy-induced state satisfying
\begin{equation}
\psi_{A_l}=0,
\qquad l=1,\ldots,N.
\label{eq:A-zero}
\end{equation}
This propagation channel is sublattice selective.  If the vacancy is instead
created on $B_1$, the imbalance mode lies on the $A$ sublattice.  Removing
$B_1$ also removes the vertical bond $B_1\!-\!A_2$, so the vacancy seed on
$A_1$ is not injected into the dimer chain and does not acquire the analogous
remote-surface component in the minimal model.

Let $\bm R=0$ denote the missing $A_1$ site, and let $\bm\delta$ be its
three intralayer displacements to $B_1$ sites. For the Fourier construction,
we restore the equation at the missing site as the auxiliary inhomogeneous
equation
\begin{align}
\gamma_0\sum_{\bm\delta}\psi_{B_1}(\bm R+\bm\delta)
&=\Lambda\delta_{\bm R,0},
\label{eq:lattice-source}\\
\gamma_0\sum_{\bm\delta}\psi_{B_l}(\bm R+\bm\delta)
+\gamma_1\psi_{B_{l-1}}(\bm R)&=0,
\quad l\geq2.
\label{eq:lattice-recursion}
\end{align}
The auxiliary amplitude $\Lambda$ fixes only the overall normalization. Choose the monolayer structure factor
\begin{equation}
g(\bm k)=1+\mathrm{e}^{-\mathrm{i}\bm k\cdot\bm a_1}
+\mathrm{e}^{-\mathrm{i}\bm k\cdot\bm a_2},
\label{eq:structure-factor}
\end{equation}
with $\bm a_1$ and $\bm a_2$ the graphene lattice vectors. Fourier transforming Eqs.~\eqref{eq:lattice-source} and \eqref{eq:lattice-recursion} yields
\begin{equation}
\widehat\psi_{B_l}(\bm k)=
\frac{\Lambda(-\gamma_1)^{l-1}}
{[\gamma_0g(\bm k)]^l},
\qquad l=1,\ldots,N,
\label{eq:lattice-wavefunction-k}
\end{equation}
and hence
\begin{equation}
\psi_{B_l}(\bm R)=
\frac{\Lambda(-\gamma_1)^{l-1}}{\sqrt{N_c}}
\sum_{\bm k}
\frac{\mathrm{e}^{\mathrm{i}\bm k\cdot\bm R}}
{[\gamma_0g(\bm k)]^l}.
\label{eq:lattice-wavefunction-r}
\end{equation}
Here $N_c$ is the number of unit cells. For a finite periodic lattice, this expression applies when the allowed
momentum mesh excludes the zeros of $g$. In a flake, the actual edge conditions
replace that definition. 

\subsection{Continuum envelope}
\label{sec:vacancy-continuum}

In the single-valley convention fixed in Sec.~\ref{sec:geometry}, the
unperturbed first-layer equation admits the monolayer vacancy solution
$\psi_{B_1}=C/\bar z$ away from the removed site
\cite{Pereira2006Disorder}. Substituting this result into
Eq.~\eqref{eq:continuum-recursion} and integrating successively gives the
particular solution,
\begin{equation}
\psi_{B_l}=
\frac{C}{(l-1)!}
\left(-\mathrm{i}\frac{\gamma_1z}{2\nu_0}\right)^{l-1}
\frac{1}{\bar z},
\qquad l=1,\ldots,N.
\label{eq:vacancy-plus}
\end{equation}
This equation is the expression quoted in the main text, where $\gamma_1/(2\nu_0)=\ell_\perp^{-1}$. The time-reversed component follows by complex conjugation and has the same radial envelope and layer weights. Writing $z=r\mathrm e^{\mathrm i\varphi}$, the radial envelope of the component retained here is independent of
$\varphi$.
With $\ell_\perp=2\nu_0/\gamma_1$, one obtains
\begin{equation}
|\psi_{B_l}(r)|=
\frac{|C|}{r}
\frac{(r/\ell_\perp)^{l-1}}{(l-1)!}.
\label{eq:radial-envelope}
\end{equation}
Thus the magnitudes in $L_1$, $L_2$, and $L_3$ scale respectively as
$r^{-1}$, $r^0$, and $r^1$.  The constant adjacent-layer magnitude for a
bilayer is the continuum limit of the extended component found for the
corresponding Bernal-vacancy class \cite{Castro2010Vacancy}.  For an outer
vacancy in an ABC trilayer, Eq.~\eqref{eq:vacancy-plus} gives a final component
$z^2/\bar z$, in agreement with the long-distance structure obtained in
Ref.~\cite{Castro2012ABC}.

\subsection{Threefold classification and the observed three-layer texture}
\label{sec:mod3}

The vacancy solution also carries a definite angular character:
\begin{equation}
\psi_{B_l}\propto
\frac{z^{l-1}}{\bar z}
=r^{l-2}\mathrm e^{\mathrm i l\varphi}.
\label{eq:vacancy-angular}
\end{equation}
Each successive layer therefore increases the winding by one. Under a rotation
by $2\pi/3$, the amplitude acquires the eigenvalue
$\exp(\!\pm2\pi\mathrm i l/3)$ up to an $l$-independent phase. Its $C_3$ representation consequently depends only on $l\bmod3$:
\begin{equation}
L_l\sim L_{l+3}\qquad\text{as $C_3$ representations}.
\label{eq:mod3-classification}
\end{equation}
We note that in the isotropic continuum theory, $|\psi_{B_l}|^2\propto r^{2l-4}$ is rotationally invariant and contains no threefold texture. The texture visible in the layer-resolved lattice densities of Fig.~1(b) of the main text and Fig.~\ref{fig:LayersN6weigth} additionally contains the atomic Bloch factors, the sublattice phases, and the three-layer recurrence of the ABC stack. The observed repetition of the threefold pattern in layers separated by three is therefore consistent with Eq.~\eqref{eq:mod3-classification} once these lattice-scale factors are included. Members of one class need not be identical: their radial envelopes differ, and finite boundaries modify the complete spatial profiles. This effect is shown in Fig.~\ref{fig:LayersN6weigth} for $N=6$. The vacancy state should also be distinguished from the impurity-induced Friedel response. A generic impurity can produce a $1/r$ asymptotic modulation of the local density of states~\cite{Dutreix2016Friedel}, whereas Eq.~\eqref{eq:vacancy-plus} describes the
sublattice-imbalance zero mode and its layer-dependent wavefunction.

\subsection{Finite-radius trends and integrated weights}
\label{sec:finite-vacancy-numerics}
\label{sec:particular-weights}

\begin{figure}
\centering
\includegraphics[width=0.95\textwidth]{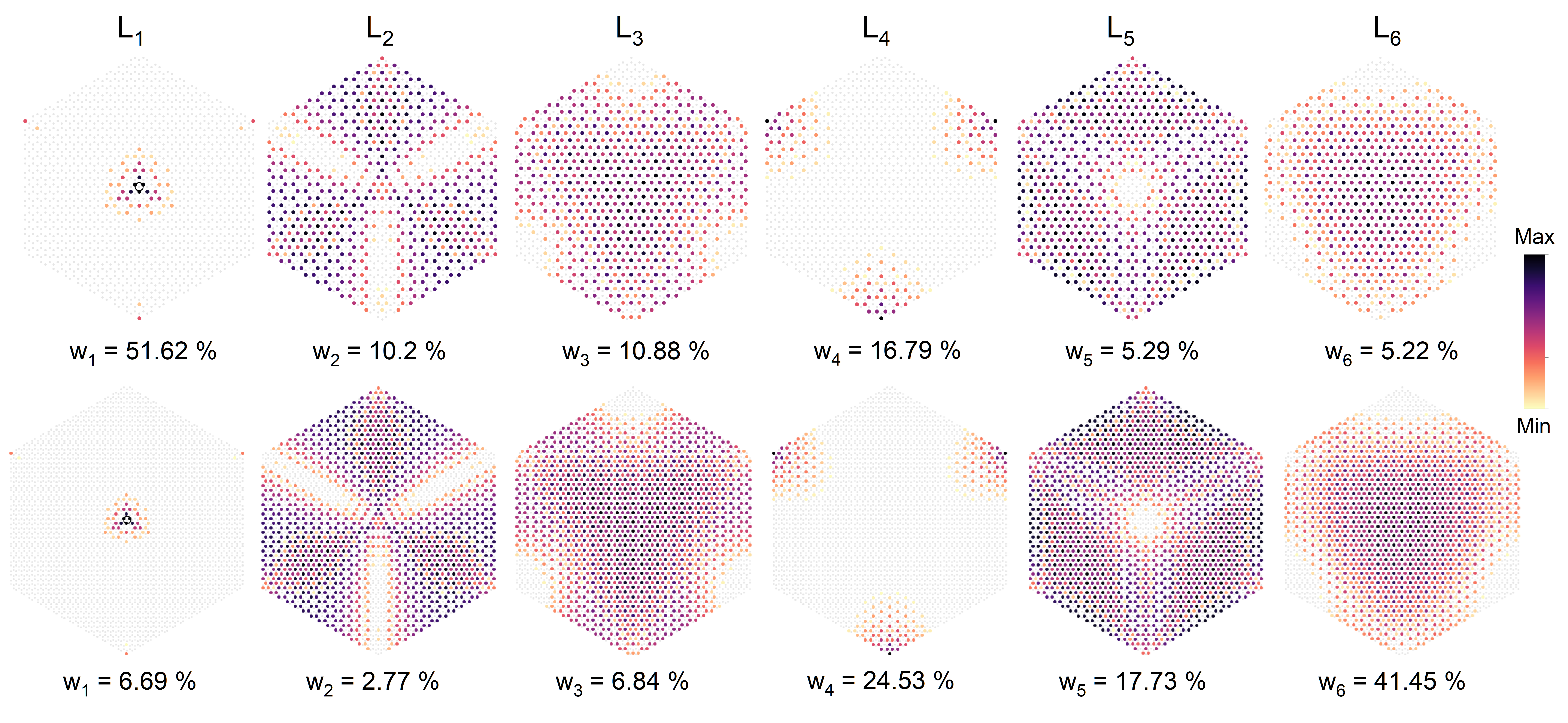}
\caption{\textbf{Size-dependent redistribution of the vacancy zero mode.}
Layer-resolved probability density in a finite, armchair-terminated $N=6$ ABC stack with an $A_1$ vacancy at the center of $L_1$. Columns correspond to $L_1,\ldots,L_6$. The top and bottom rows show $R=39.76$~\AA ($X=1.06$) and
$R=61.06$~\AA ($X=1.63$), respectively.  The parameters are $a_{cc}=1.42$~\AA, $\gamma_0=3.16$~eV, and $\gamma_1=0.36$~eV.
Each panel is independently color normalized to emphasize its spatial
profile; the percentages below the panels give the integrated layer weights
defined in Eq.~\eqref{eq:finite-vacancy-weight}. Increasing the lateral size reduces the vacancy-layer weight from $51.6\%$ to $6.7\%$ and increases the remote-layer weight from
$5.2\%$ to $41.5\%$, making $L_6$ the dominant layer. The redistribution is nonmonotonic across the stack, with $L_4$ retaining a substantial contribution of $24.5\%$ in the larger flake. The spatial profiles also exhibit the three-layer recurrence discussed in the main text, particularly between $L_2$ and $L_5$, and between $L_3$ and $L_6$, despite their different integrated weights.}
\label{fig:LayersN6weigth}
\end{figure}

The vacancy solution is not square integrable on the infinite plane, so its
normalized layer distribution must be defined in a finite sample. To illustrate the effects we choose a flake with an armchair termination because, unlike an ideal zigzag edge, it does not support a sublattice-polarized zero-energy mode. This avoids an independent family of edge zero modes and permits the vacancy-induced state to be followed without ambiguity. For these flakes, an integer $n_f$ determines the nominal radial scale $R=a_{cc}(n_f-1)$. We define
\begin{equation}
\ell_\perp=\frac{2\nu_0}{\gamma_1}
=\frac{3a_{cc}\gamma_0}{\gamma_1},
\qquad
X=\frac{R}{\ell_\perp}
=\frac{R\gamma_1}{3a_{cc}\gamma_0},
\label{eq:length-X}
\end{equation}
with $a=\sqrt{3}a_{cc}$. We integrate the continuum envelope over a circular annulus to obtain the size-dependent trends. Quantitative weights for the open armchair flake are obtained independently from its lattice eigenstate below. With a core cutoff $r_0$, the radial integral gives, up to a constant factor,
\begin{align}
w_1&=2\pi|C|^2\ln(R/r_0),
\label{eq:w1}\\
w_l&=\frac{2\pi|C|^2}{2l-2}
\frac{X^{2l-2}-X_0^{2l-2}}{[(l-1)!]^2},
\qquad l\geq2.
\label{eq:wl}
\end{align}
Here $X_0=r_0/\ell_\perp$. Neglecting the core power for $R\gg r_0$ and
suppressing the common angular factor $2\pi$ gives the equivalent leading
expressions quoted in the main text,
\begin{equation}
\begin{aligned}
w_1&\simeq |C|^2\ln(R/r_0),\\
w_l&\simeq
\frac{|C|^2X^{2l-2}}{(2l-2)[(l-1)!]^2},
\qquad l\geq2.
\end{aligned}
\label{eq:vacancy-weights}
\end{equation}
The common factor cancels upon normalization and does not alter any of the ratios below. Every term of this particular solution diverges as $R\rightarrow\infty$: logarithmically in the vacancy layer and as $R^{2l-2}$ in later layers.  The zero mode is therefore not normalizable on the infinite plane.  For fixed finite $N$, the highest power in Eq.~\eqref{eq:wl} places the normalized large-$R$ weight on $L_N$, with the probability pushed toward the outer radial boundary.  The unnormalized large-$R$ behavior is thus distinct from a normalizable infinite-plane state. For this particular solution, the pointwise ratios are
\begin{equation}
\frac{|\psi_{B_{l+1}}(r)|}{|\psi_{B_l}(r)|}
=\frac{r}{l\ell_\perp},
\qquad
\frac{|\psi_{B_N}(R)|^2}{|\psi_{B_1}(R)|^2}
=\left[\frac{X^{N-1}}{(N-1)!}\right]^2,
\label{eq:vacancy-ratios}
\end{equation}
as quoted in the main text.
At fixed $R$, the adjacent boundary amplitudes obey
\begin{equation}
\frac{|\psi_{B_{l+1}}(R)|}{|\psi_{B_l}(R)|}=\frac{X}{l}.
\label{eq:layer-ratio-R}
\end{equation}
The integrated probabilities obey a different ratio.  For
$R\gg r_0$ and $l\geq2$,
\begin{equation}
\frac{w_{l+1}}{w_l}\simeq X^2\frac{l-1}{l^3}.
\label{eq:integrated-layer-ratio}
\end{equation}
Thus the integrated weight continues to increase from $L_l$ to $L_{l+1}$ only
while $X^2(l-1)>l^3$. For this particular solution, all integrated weights from $L_2$ through
$L_N$ increase successively only if
\begin{equation}
X>X_N^{(w)}\equiv
\sqrt{\frac{(N-1)^3}{N-2}},
\qquad N\geq3,
\label{eq:physical-crossover}
\end{equation}
because $l^3/(l-1)$ increases for $l\geq2$. This condition does not include
$L_1$, and it is not a threshold for the onset of an interior maximum.
In physical variables,
\begin{equation}
R>R_N^{(w)}=
\frac{3a_{cc}\gamma_0}{\gamma_1}
\sqrt{\frac{(N-1)^3}{N-2}}.
\label{eq:physical-radius}
\end{equation}
For the parameters in Table~\ref{tab:finite-vacancy-weights} (see also Fig.~\ref{fig:LayersN6weigth}), 
$\ell_\perp=37.39$~\AA. Thus $X>5.59$, or $R>20.9$~nm, is required
for the particular weights to increase successively from $L_2$ to $L_6$.

The momentum representation gives complementary intuition. A local defect
samples long-wavelength momenta around the selected Dirac point, and a sample of radius
$R$ introduces an infrared scale of order $R^{-1}$. In a translation-invariant
bulk channel the layer-step magnitude is $\gamma_1/(\nu_0|\bm p|)$.
Increasing $R$ therefore permits smaller momenta with stronger relative
remote-layer amplitudes. In an open flake, however, boundary matching fixes
their combination and the homogeneous terms. 

\begin{figure}
\centering
\includegraphics[width=0.40\textwidth]{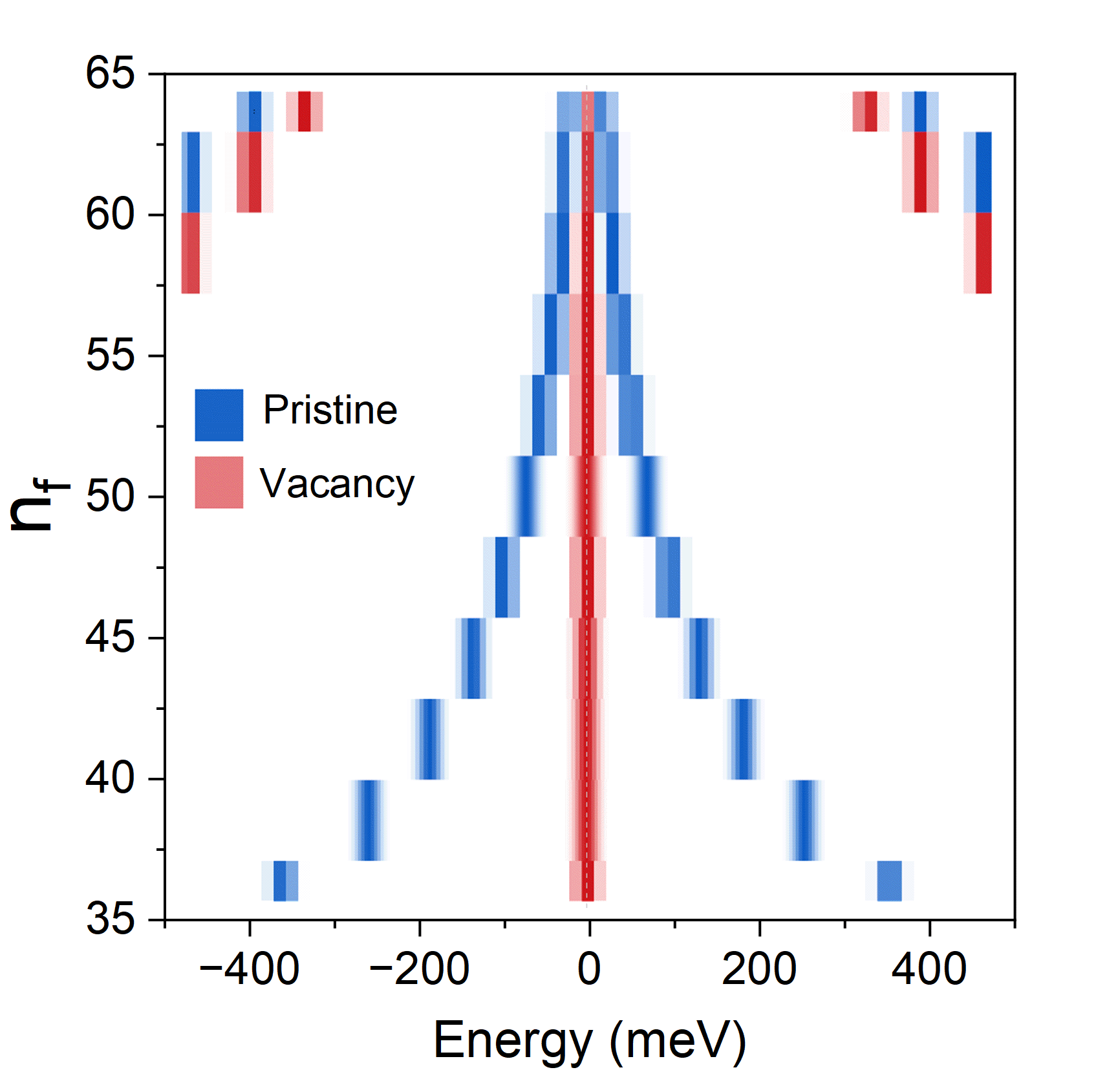}
\caption{\textbf{Size evolution of the density of states.} Total density of states of pristine (blue) and $A_1$-vacancy (red) armchair flakes of rhombohedral graphene, as a function of energy and the size index $n_f$. Increasing $n_f$ increases the lateral extent of the flake.  Only sizes for which the pristine flake has no zero modes are included; the discrete sampling of these selected sizes produces the sharp changes along the $n_f$ axis. The pristine system exhibits a pair of surface-derived levels at opposite energies whose splitting decreases as the flake grows. In contrast, the vacancy produces a zero mode pinned at $E=0$ throughout the size range shown, as required by the sublattice imbalance in the minimal chiral model. Additional finite-energy levels enter the displayed window at larger sizes and are shifted by the vacancy. The central peaks illustrate the distinction between the finite-size splitting of the pristine surface states and the vacancy-induced zero mode.}
\label{fig:DOSFinite}
\end{figure}
\subsection{Finite-lattice systems}
\label{sec:finite-lattice-state}

We now evaluate the vacancy-induced redistribution in a finite real-space rhombohedral multilayer system. Our arguments are valid for any size, but to illustrate the effects, we use six equal-sized, armchair-terminated graphene flakes with open in-plane boundaries, shown in Fig.~\ref{fig:LayersN6weigth} for $N=6$. A vacancy is introduced at the center of $L_1$ by removing one $A_1$ site and all incident hopping matrix elements. 

For each flake radius $R$, we numerically diagonalize both the pristine and vacancy systems. As shown in Fig.~\ref{fig:DOSFinite}, the pristine finite stack contains the two usual surface low-energy states at approximately $\pm E_{\mathrm s}(R)$ depending on the system size; their finite splitting decreases as the flake grows. Removing one $A_1$ site guarantees an exact zero-energy state through the sublattice imbalance. To isolate that state from
accidental boundary levels, we retain only sizes for which the pristine flake
has no eigenstate inside a narrow selection window
$|E|\leq10^{-5}~\mathrm{eV}$, while the vacancy flake has exactly one. The
selected zero-energy eigenstate is the finite-system counterpart of the
vacancy mode. It obeys the lattice zero-energy equations and, in the continuum bulk limit, Eq.~\eqref{eq:continuum-recursion}. It need not equal the particular field in Eq.~\eqref{eq:vacancy-plus}; the armchair boundary fixes its complete global form.

\begin{table}[t]
\caption{Layer-resolved weight of the vacancy state in a finite $N=6$ ABC stack. The vacancy is located on $A_1$ in $L_1$. The parameters are
$a_{cc}=1.42~\text{\AA}$,
$\gamma_0=3.16~\mathrm{eV}$, and $\gamma_1=0.36~\mathrm{eV}$, giving
$\ell_\perp=3a_{cc}\gamma_0/\gamma_1$. Only flake sizes with no pristine eigenstate but exactly one vacancy-induced eigenstate inside the selection window $|E|\leq10^{-5}~\mathrm{eV}$ are included.}
\centering
\small
\begin{tabular}{cc*{6}{r}}
\hline\hline
$R$ (\AA) & $X$ & \multicolumn{6}{c}{$W_l^{\mathrm{vac}}$ (\%)} \\
\cline{3-8}
 & & $L_1$ & $L_2$ & $L_3$ & $L_4$ & $L_5$ & $L_6$ \\
\hline
14.20 & 0.3797 & 95.64 &  3.66 &  0.57 &  0.12 &  0.01 &  0.00 \\
18.46 & 0.4937 & 92.91 &  5.27 &  1.31 &  0.46 &  0.03 &  0.01 \\
22.72 & 0.6076 & 88.94 &  6.99 &  2.55 &  1.33 &  0.15 &  0.05 \\
26.98 & 0.7215 & 83.18 &  8.62 &  4.35 &  3.16 &  0.48 &  0.22 \\
31.24 & 0.8354 & 75.06 &  9.90 &  6.62 &  6.38 &  1.27 &  0.77 \\
35.50 & 0.9494 & 64.36 & 10.50 &  8.99 & 11.11 &  2.82 &  2.22 \\
39.76 & 1.0633 & 51.62 & 10.20 & 10.88 & 16.79 &  5.29 &  5.22 \\
44.02 & 1.1772 & 38.40 &  9.02 & 11.73 & 22.11 &  8.48 & 10.26 \\
48.28 & 1.2911 & 26.56 &  7.31 & 11.39 & 25.74 & 11.80 & 17.20 \\
52.54 & 1.4051 & 17.33 &  5.52 & 10.15 & 27.09 & 14.62 & 25.28 \\
56.80 & 1.5190 & 10.88 &  3.97 &  8.50 & 26.45 & 16.61 & 33.58 \\
61.06 & 1.6329 &  6.69 &  2.77 &  6.84 & 24.53 & 17.73 & 41.45 \\
65.32 & 1.7468 &  4.08 &  1.90 &  5.36 & 21.99 & 18.13 & 48.53 \\
\hline\hline
\end{tabular}
\label{tab:finite-vacancy-weights}
\end{table}

From the normalized lattice wavefunction $\Psi^{\mathrm{vac}}$, we define the
weight in layer $L_l$ as
\begin{equation}
W_l^{\mathrm{vac}}=
\sum_{\bm R\in L_l}\sum_{\alpha=A,B}
\left|\Psi^{\mathrm{vac}}_{l\alpha}(\bm R)\right|^2,
\qquad
\sum_{l=1}^{N}W_l^{\mathrm{vac}}=1.
\label{eq:finite-vacancy-weight}
\end{equation}
This is the quantity written as $W_l$ in the main text. In Tab.~\ref{tab:finite-vacancy-weights} we show $W_l^{\mathrm{vac}}$ for different system sizes. We characterize the system size by the radial boundary scale $R$ and use $X=R/\ell_\perp$ defined in Eq.~\eqref{eq:length-X}.

The values in Table~\ref{tab:finite-vacancy-weights} give a direct
finite-lattice test of the size-dependent redistribution discussed in the
main text. For $X<1$, the zero mode is concentrated predominantly near the
vacancy surface. At the smallest radius, $95.64\%$ of its normalized weight
lies in $L_1$, while the remote-layer contribution is negligible. As the
radius increases, $W_1^{\mathrm{vac}}$ decreases and the state develops
substantial weight throughout the stack. The scale $X\sim1$ therefore marks
the crossover to appreciable propagation away from the vacancy surface in
this finite geometry. For larger flakes the transfer toward the opposite surface becomes
pronounced. At $X=1.4051$, the remote-layer contribution,
$W_6^{\mathrm{vac}}=25.28\%$, already exceeds the vacancy-layer contribution,
$W_1^{\mathrm{vac}}=17.33\%$, although $L_4$ carries the largest individual
weight. At $X=1.7468$, the remote layer is clearly dominant, with
$W_6^{\mathrm{vac}}=48.53\%$, while only $4.08\%$ remains in $L_1$.

We note that the redistribution need not be monotonic in layer index: intermediate layers can carry substantial or even maximal weight. The continuum radial powers describe the size-driven trend, while Table~\ref{tab:finite-vacancy-weights} determines the quantitative layer distribution in the chosen finite geometry. 

\section{Circular void in the first layer}
\label{sec:void}

\subsection{Geometry, angular channels and the circular scar}
\label{sec:void-seed}

We now consider a circular void of radius $R_0$ in $L_1$. Without loss of generality and to obtain analytical expressions we use an ideal continuum zigzag boundary condition that supports the $B_1$ sector. In the single-valley convention of Sec.~\ref{sec:geometry}, the exterior first-layer equation is $\partial_z\psi_{B_1}=0$. Its decaying, single-valued angular channels are
\begin{equation}
\psi_{B_1}^{(m)}(z,\bar z)=\frac{C_m}{\bar z^m},
\qquad r>R_0,\qquad m=1,2,\ldots.
\label{eq:void-seed}
\end{equation}
The amplitude is absent inside the first-layer void. The index $m$ labels the circular boundary channel equivalent to the continuous longitudinal momentum of a straight edge.

\begin{figure}
\centering
\includegraphics[width=0.99\textwidth]{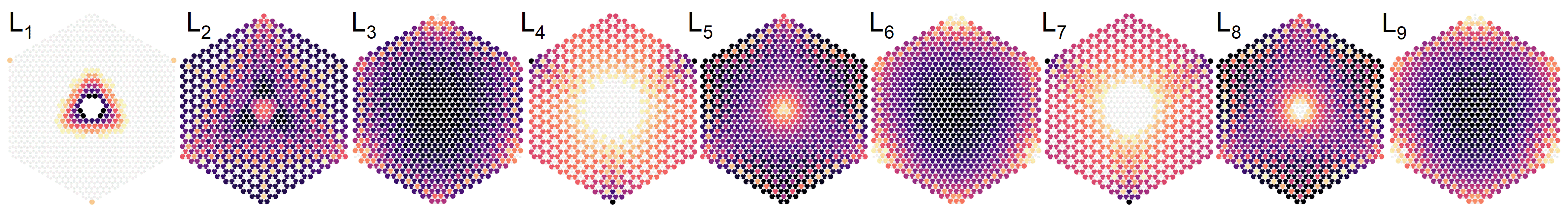}
\caption{\textbf{Layer-resolved structure of a void-induced zero mode in an $N=9$ ABC stack.} Site-resolved probability density of the selected $B$-polarized zero mode across layers $L_1,\ldots,L_9$. The physical void is introduced only in $L_1$; all subsequent layers contain the complete graphene lattice. Nevertheless, the propagated wavefunction retains a pronounced spatial memory of the void, producing density depletions and ring-like structures in later layers. The profiles also display the characteristic three-layer recurrence of the ABC stacking. The central depletion visible in the later layers is therefore a wavefunction scar inherited through the layer-by-layer propagation, rather than a physical hole in those layers. Each panel is independently color normalized to emphasize its spatial
profile.}
\label{fig:SMVoidN9}
\end{figure}

For $l\geq2$, Eq.~\eqref{eq:continuum-recursion} gives
\begin{equation}
\partial_z\psi_{B_l}^{(m)}=\mu\psi_{B_{l-1}}^{(m)},
\qquad
\mu=-\mathrm{i}\frac{\gamma_1}{2\nu_0}.
\label{eq:void-recursion}
\end{equation}
To construct the component that retains the void's scar, take the lower-layer amplitudes to vanish for $r<R_0$ and match continuously to zero at $r=R_0$. Introduce
\begin{equation}
\chi(z,\bar z)=z-\frac{R_0^2}{\bar z},
\qquad
\partial_z\chi=1,
\qquad
\chi\big|_{r=R_0}=0.
\label{eq:void-chi}
\end{equation}
Since $\partial_z\bar z^{-m}=0$ on the exterior domain,
\begin{equation}
\partial_z\left[
\frac{\chi^{l-1}}{(l-1)!\bar z^m}
\right]
=
\frac{\chi^{l-2}}{(l-2)!\bar z^m}.
\label{eq:void-identity}
\end{equation}
Successive integration then gives
\begin{equation}
\psi_{B_l}^{(m)}(z,\bar z)=
\begin{cases}
0,&r\leq R_0,\\[2pt]
\displaystyle
\frac{C_m\mu^{l-1}}{(l-1)!}
\frac{\left(z-R_0^2/\bar z\right)^{l-1}}{\bar z^m},
&r>R_0,
\end{cases}
\qquad l=1,\ldots,N.
\label{eq:void-solution}
\end{equation}
In particular, the first two propagated components for $r>R_0$ are
\begin{align}
\psi_{B_2}^{(m)}
&=
C_m\mu\left[
\frac{z}{\bar z^m}
-\frac{R_0^2}{\bar z^{m+1}}
\right],
\\
\psi_{B_3}^{(m)}
&=
\frac{C_m\mu^2}{2}
\left[
\frac{z^2}{\bar z^m}
-\frac{2R_0^2z}{\bar z^{m+1}}
+\frac{R_0^4}{\bar z^{m+2}}
\right].
\label{eq:void-first-layers}
\end{align}
The compact power in Eq.~\eqref{eq:void-solution} contains all terms generated by the successive integrations. For every $l\geq2$, the exterior amplitude vanishes at the circle and its zero extension introduces no delta-function source in the first-order layer equation. The density depletion beneath the void is retained in this propagated component, even though the underlying layers contain no physical hole. The void can therefore leave a spatial imprint, or scar, on the remote surface. Equation~\eqref{eq:void-solution}
describes this scar-preserving channel; the complete, non-ideal zero mode may also contain additional components that modify the depletion beneath the void. A finite-lattice example of a void-induced state is shown in Fig.~\ref{fig:SMVoidN9}.

\subsection{Radial trends and layer weights}
\label{sec:void-weights}

Writing $z=r\mathrm e^{\mathrm i\varphi}$ gives $\chi=(r-R_0^2/r)\mathrm e^{\mathrm i\varphi}$. For positive $\gamma_1$, $\mu=-\mathrm i/\ell_\perp$, and the amplitude has angular winding $m+l-1$. Its envelope and adjacent-layer ratio are
\begin{align}
|\psi_{B_l}^{(m)}(r)|
&=
\frac{|C_m|}{(l-1)!r^m}
\left(
\frac{r^2-R_0^2}{\ell_\perp r}
\right)^{l-1},
\label{eq:void-envelope}
\\
\frac{|\psi_{B_{l+1}}^{(m)}(r)|}
{|\psi_{B_l}^{(m)}(r)|}
&=
\frac{r^2-R_0^2}{l\ell_\perp r},
\qquad r>R_0.
\label{eq:void-layer-ratio}
\end{align}
Near the circle, the propagated magnitude vanishes as $(r-R_0)^{l-1}$; far away, it scales as $r^{l-1-m}$. At a fixed radius the ratio in Eq.~\eqref{eq:void-layer-ratio} is independent of $m$, but $m$ determines which radius contribute most strongly to the norm. Smaller $m$ gives a slower radial decay and makes the weight more sensitive to the available lateral extent.

For a radial cutoff $R>R_0$, set $\rho=r/R_0$, $\rho_{\max}=R/R_0$, and $Y=R_0/\ell_\perp$. The integrated weights of the channels are
\begin{equation}
w_l^{(m)}(R)=
\frac{
2\pi|C_m|^2R_0^{2-2m}Y^{2l-2}
}{
[(l-1)!]^2
}
\int_1^{\rho_{\max}}
\rho^{1-2m}
\left(\rho-\rho^{-1}\right)^{2l-2}
\,\mathrm d\rho,
\qquad
W_l^{(m)}
=
\frac{w_l^{(m)}}{\sum_{j=1}^{N}w_j^{(m)}}.
\label{eq:void-finite-weights}
\end{equation}
This expression displays the dependence on $m$, $Y$, and the outer cutoff $R/R_0$ without identifying an annular integral with an exact finite-flake eigenstate. On the unbounded exterior domain, the entire channel is square-integrable only for $m>N$; the $m=N$ remote-layer norm diverges logarithmically. For $m>N$, the substitution $t=\rho^{-2}$ gives
\begin{equation}
w_l^{(m)}(\infty)=
\frac{
\pi|C_m|^2R_0^{2-2m}Y^{2l-2}
}{
[(l-1)!]^2
}
\mathrm B(m-l,2l-1),
\label{eq:void-infinite-weights}
\end{equation}
where $\mathrm B(a,b)$ is the Euler beta function. Consequently,
\begin{equation}
\frac{w_{l+1}^{(m)}}{w_l^{(m)}}
=
\frac{
2(2l-1)Y^2
}{
l(m-l-1)(m+l-1)
},
\qquad
l<N,\quad m>N.
\label{eq:void-integrated-ratio}
\end{equation}
For the normalizable channels $m>N$, the layer distribution is controlled by both the void radius and the angular structure of the boundary state. Increasing $R_0/\ell_\perp$ enhances the relative weight on successive layers, while increasing $m$ suppresses it. Thus, a larger void or a lower angular channel can favor the propagation of the void-induced state toward the remote surface. For $m\leq N$, the channel requires a finite outer scale for normalization, and its layer distribution also depends on that scale.
These results connect the spatial structure of the boundary state to the strength of its remote-surface imprint.

At the first-layer level, conformal maps can generate seeds associated with other void shapes. Their continuation through the multilayer must then be obtained from the original layer equations in the physical coordinates~\cite{katsnelson_transport_2008,rycerz_conformal_2009}.

\subsection{Finite void construction and selection of geometries}
\label{sec:void-finite-construction}

For the finite-lattice calculation, we use the same armchair-terminated ABC flakes introduced in the previous section. The pristine stack is constructed first, and the void is then cut exclusively from $L_1$ by removing the $A_1$ and $B_1$ sites lying inside a given boundary. Layers $L_2,\ldots,L_N$ are left unchanged. The lattice void is parametrized by a circumscribed radius $R_0$ and a roundness parameter $f$. The limit $f=1$ corresponds to a circular cut, whereas decreasing $f$ toward $1/2$ continuously deforms the boundary into a sharper triangular zigzag termination. In the calculations reported here we use $f=0.85$, which produces a rounded triangular void while preserving the sublattice character of the zigzag boundary. Its orientation is chosen such that the associated zero mode is supported on the $B$ sublattice. The number of zero-energy states depends jointly on the void radius, the flake size, the number of layers, and the roundness parameter. In particular, varying $f$ can generate additional zero modes associated with the detailed atomic structure of the boundary. Since our purpose here is to isolate the layer-by-layer charge-transfer mechanism, we select combinations of $(R_0,n_f,N,f)$ for which the void supports a single zero mode, as summarized in Table~\ref{tab:finite-void-weights}. This avoids additional boundary modes that would otherwise obscure the redistribution of the selected state across the rhombohedral stack.

Although the analytical solutions above are derived for an ideal circular void, their essential propagation mechanism is not restricted to a perfectly circular boundary. Deforming the boundary modifies the detailed spatial and angular structure of the seed while preserving the sublattice-polarized channel that is subsequently continued through the ABC stack. The circular solution therefore provides the analytical reference for the layer redistribution and normalization trends observed in the finite lattice  geometry. We use the same definitions of the flake radius $R$, dimensionless size $X=R/\ell_\perp$, and layer-resolved weights introduced for the vacancy calculation. For the void, the additional dimensionless scale is
\begin{equation}
X_0=\frac{R_0}{\ell_\perp},
\label{eq:void-X0}
\end{equation}
where $R_0$ is the void radius. The layer weights $W_l^{\mathrm{void}}$ are obtained from the normalized void eigenstate using the same layer projection as in Eq.~\eqref{eq:finite-vacancy-weight}.

\begin{table}[t]
\caption{\textbf{Layer-resolved weight of the void-induced state in a finite $N=6$ ABC stack.} The void is introduced only in $L_1$ as a rounded triangular zigzag termination of circumscribed radius $R_0$, with roundness parameter $f=0.85$ and orientation chosen such that the selected zero mode is $B$ polarized. The parameters are $a_{cc}=1.42~\text{\AA}$, $\gamma_0=3.16~\mathrm{eV}$, and $\gamma_1=0.36~\mathrm{eV}$. Every configuration listed has no pristine eigenstate and exactly one void-induced eigenstate inside the selection window.}
\centering
\small
\begin{tabular}{cc*{6}{r}}
\hline\hline
$R$ (\AA) & $X$ & \multicolumn{6}{c}{$W_l^{\mathrm{void}}$ (\%)} \\
\cline{3-8}
 & & $L_1$ & $L_2$ & $L_3$ & $L_4$ & $L_5$ & $L_6$ \\
\hline
\multicolumn{8}{c}{$R_0=3~\text{\AA}$, \quad $X_0=0.0802$} \\
\hline
14.20 & 0.3797 & 89.76 &  8.92 &  1.14 &  0.15 &  0.02 &  0 \\
18.46 & 0.4937 & 84.88 & 12.08 &  2.40 &  0.51 &  0.11 &  0.02 \\
22.72 & 0.6076 & 78.79 & 15.07 &  4.30 &  1.31 &  0.40 &  0.12 \\
26.98 & 0.7215 & 71.18 & 17.51 &  6.80 &  2.82 &  1.19 &  0.50 \\
31.24 & 0.8354 & 61.80 & 18.93 &  9.59 &  5.22 &  2.87 &  1.59 \\
35.50 & 0.9494 & 50.73 & 18.89 & 12.10 &  8.36 &  5.84 &  4.09 \\
39.76 & 1.0633 & 38.79 & 17.21 & 13.61 & 11.63 & 10.05 &  8.71 \\
44.02 & 1.1772 & 27.47 & 14.29 & 13.67 & 14.16 & 14.83 & 15.58 \\
48.28 & 1.2911 & 18.14 & 10.92 & 12.44 & 15.34 & 19.17 & 23.99 \\
52.54 & 1.4051 & 11.38 &  7.84 & 10.48 & 15.19 & 22.30 & 32.81 \\
56.80 & 1.5190 &  6.92 &  5.41 &  8.39 & 14.12 & 24.06 & 41.10 \\
61.06 & 1.6329 &  4.16 &  3.66 &  6.51 & 12.59 & 24.66 & 48.42 \\
65.32 & 1.7468 &  2.51 &  2.46 &  4.98 & 10.96 & 24.44 & 54.66 \\
\hline
\multicolumn{8}{c}{$R_0=6~\text{\AA}$, \quad $X_0=0.1605$} \\
\hline
14.20 & 0.3797 & 94.65 &  4.57 &  0.64 &  0.13 &  0.01 &  0 \\
18.46 & 0.4937 & 90.76 &  6.89 &  1.71 &  0.59 &  0.05 &  0.01 \\
22.72 & 0.6076 & 85.21 &  9.23 &  3.49 &  1.81 &  0.20 &  0.06 \\
26.98 & 0.7215 & 77.55 & 11.26 &  5.94 &  4.30 &  0.65 &  0.30 \\
31.24 & 0.8354 & 67.45 & 12.58 &  8.78 &  8.46 &  1.69 &  1.03 \\
35.50 & 0.9494 & 55.23 & 12.80 & 11.41 & 14.12 &  3.60 &  2.84 \\
39.76 & 1.0633 & 42.10 & 11.84 & 13.10 & 20.24 &  6.40 &  6.33 \\
44.02 & 1.1772 & 29.81 &  9.97 & 13.41 & 25.29 &  9.72 & 11.80 \\
48.28 & 1.2911 & 19.80 &  7.75 & 12.46 & 28.17 & 12.93 & 18.89 \\
52.54 & 1.4051 & 12.55 &  5.68 & 10.74 & 28.68 & 15.50 & 26.84 \\
56.80 & 1.5190 &  7.74 &  4.00 &  8.80 & 27.39 & 17.22 & 34.86 \\
61.06 & 1.6329 &  4.71 &  2.76 &  6.97 & 25.04 & 18.12 & 42.40 \\
65.32 & 1.7468 &  2.86 &  1.88 &  5.42 & 22.25 & 18.37 & 49.21 \\
\hline
\multicolumn{8}{c}{$R_0=8~\text{\AA}$, \quad $X_0=0.2139$} \\
\hline
14.20 & 0.3797 & 92.61 &  6.49 &  0.79 &  0.10 &  0.01 &  0 \\
18.46 & 0.4937 & 86.85 & 10.50 &  2.10 &  0.44 &  0.09 &  0.02 \\
22.72 & 0.6076 & 79.92 & 14.17 &  4.13 &  1.26 &  0.39 &  0.12 \\
26.98 & 0.7215 & 71.48 & 17.13 &  6.83 &  2.85 &  1.20 &  0.50 \\
31.24 & 0.8354 & 61.32 & 18.88 &  9.83 &  5.38 &  2.96 &  1.63 \\
35.50 & 0.9494 & 49.64 & 18.95 & 12.48 &  8.65 &  6.05 &  4.23 \\
39.76 & 1.0633 & 37.38 & 17.25 & 14.00 & 12.00 & 10.38 &  8.99 \\
44.02 & 1.1772 & 26.07 & 14.25 & 13.97 & 14.51 & 15.21 & 15.98 \\
48.28 & 1.2911 & 16.99 & 10.84 & 12.62 & 15.61 & 19.51 & 24.43 \\
52.54 & 1.4051 & 10.55 &  7.75 & 10.57 & 15.37 & 22.56 & 33.20 \\
56.80 & 1.5190 &  6.38 &  5.33 &  8.43 & 14.22 & 24.24 & 41.41 \\
61.06 & 1.6329 &  3.82 &  3.60 &  6.53 & 12.64 & 24.77 & 48.64 \\
65.32 & 1.7468 &  2.29 &  2.42 &  4.98 & 10.99 & 24.51 & 54.81 \\
\hline\hline
\end{tabular}
\label{tab:finite-void-weights}
\end{table}

The finite-lattice results reproduce the central trend of the continuum solution. For a fixed void, increasing the lateral extent of the flake progressively transfers normalized weight away from the perturbed surface and toward the later layers of the ABC stack. For example, for $R_0=3~\text{\AA}$, the $L_1$ contribution decreases from $89.76\%$ at $X=0.3797$ to $2.51\%$ at $X=1.7468$, while the remote-layer contribution increases from essentially zero to $54.66\%$. The same redistribution occurs for the other void sizes: at $X=1.7468$, $L_6$ carries $49.21\%$ and $54.81\%$ of the state for $R_0=6~\text{\AA}$ and $8~\text{\AA}$, respectively. The transfer toward the remote surface is therefore not tied to a particular choice of the void radius and is a general tendency.

The lattice boundary nevertheless affects how the weight is distributed among the intermediate layers. This is most evident for $R_0=6~\text{\AA}$, where at $X=1.4051$ the largest contribution occurs on $L_4$, with $W_4^{\mathrm{void}}=28.68\%$, slightly exceeding $W_6^{\mathrm{void}}=26.84\%$. Upon further increasing the flake size, the remote layer becomes dominant, reaching $W_6^{\mathrm{void}}=34.86\%$ at $X=1.5190$ and $49.21\%$ at $X=1.7468$. Thus the detailed layer partition depends on the microscopic boundary realization, while the global transfer of normalized weight from the void layer toward the opposite surface remains common to all three cases. This distinction is consistent with the continuum picture: the boundary determines the seed and its angular content, whereas the subsequent layer-by-layer propagation is fixed by the ABC stacking recursion.

\section{Zigzag termination of the first layer}
\label{sec:zigzag-step}

\subsection{Continuum propagation from a zigzag termination}
\label{sec:zigzag-continuum}

We now consider a geometry in which $L_2,\ldots,L_N$ extend over the full
plane, whereas $L_1$ occupies the half-plane $y>0$ and terminates at a
straight zigzag step parallel to $x$. For fixed ABC stack and edge
direction, there are two inequivalent terminations of $L_1$, exposing either
the $A_1$ or the $B_1$ sublattice~\cite{Fujita1996Peculiar}. In our stacking
convention, $A_1$ is undimerized, whereas $B_1$ is vertically coupled to $A_2$. This distinction determines whether the zigzag zero mode enters the $B_1\rightarrow B_N$ propagation channel.

Because the system is translationally invariant along the zigzag edge,
each boundary mode has a conserved momentum along $x$. Let $\kappa>0$ denote the magnitude of this momentum measured from the projected Dirac point. At zero energy, the Dirac equation gives a boundary wavefunction that decays into $y>0$ as $\mathrm{e}^{-\kappa y}$. The penetration length of its amplitude is therefore $\kappa^{-1}$, while its probability density decays as $\mathrm{e}^{-2\kappa y}$ and has a penetration length $(2\kappa)^{-1}$.

An $A_1$-terminated edge supports the usual sublattice-polarized solution on
the undimerized $A_1$ orbital. In the single-valley convention of
Sec.~\ref{sec:geometry}, it may be written as
\begin{equation}
\psi_{A_1}(x,y;\kappa)
=
\widetilde C_\kappa
\mathrm{e}^{\mathrm{i}\kappa x-\kappa y},
\qquad
\psi_{A_l}(x,y;\kappa)=0,
\quad l=2,\ldots,N.
\label{eq:zigzag-A-confined}
\end{equation}
Because $A_1$ has no vertical partner in the minimal model, this solution
remains confined to $L_1$. A $B_1$-terminated edge supports the complementary solution and directly seeds the propagating channel. Its first-layer profile can be chosen as
\begin{equation}
\psi_{B_1}(x,y;\kappa)
=
C_\kappa
\mathrm{e}^{-\mathrm{i}\kappa x-\kappa y},
\qquad
\kappa>0.
\label{eq:zigzag-seed}
\end{equation}
Here $\pi^\dagger=-\mathrm{i}\partial_x-\partial_y$, and therefore
$\pi^\dagger\psi_{B_1}=0$. Writing
\begin{equation}
\psi_{B_l}(x,y;\kappa)
=
C_\kappa F_l(y)
\mathrm{e}^{-\mathrm{i}\kappa x-\kappa y},
\qquad
F_1(y)=1,
\label{eq:zigzag-Fl-ansatz}
\end{equation}
and substituting into Eq.~\eqref{eq:continuum-recursion} gives
\begin{equation}
\frac{\mathrm{d}F_l}{\mathrm{d}y}
=
\frac{\gamma_1}{\nu_0}F_{l-1},
\qquad
l=2,\ldots,N.
\label{eq:zigzag-Fl-recursion}
\end{equation}
The layers $L_2,\ldots,L_N$ have no physical edge and contain no independent
boundary source. To isolate the component generated by the $B_1$ termination,
we impose $F_{l\geq2}(0)=0$. Successive integration
then gives
\begin{equation}
F_l(y)
=
\frac{1}{(l-1)!}
\left(\frac{\gamma_1y}{\nu_0}\right)^{l-1},
\end{equation}
and hence
\begin{equation}
\psi_{B_l}(x,y;\kappa)
=
\frac{C_\kappa}{(l-1)!}
\left(\frac{\gamma_1y}{\nu_0}\right)^{l-1}
\mathrm{e}^{-\mathrm{i}\kappa x-\kappa y},
\qquad
l=1,\ldots,N,
\label{eq:zigzag-propagated}
\end{equation}
for $y>0$, up to convention-dependent layer phases. The components in $L_2,\ldots,L_N$ are generated sequentially by the ABC recursion. Every layer retains the exponential decay of the initial boundary mode, while the additional power of $y$ moves its density maximum progressively farther from the termination.

The norm per unit edge length obtained from
Eq.~\eqref{eq:zigzag-propagated} is finite for $\kappa>0$. The unnormalized
weight on $L_l$ is
\begin{equation}
w_l(\kappa)
=
\frac{|C_\kappa|^2}{[(l-1)!]^2}
\left(\frac{\gamma_1}{\nu_0}\right)^{2l-2}
\int_0^\infty
y^{2l-2}\mathrm{e}^{-2\kappa y}\,\mathrm{d}y .
\label{eq:zigzag-continuum-weight}
\end{equation}
Using the interlayer length
$\ell_\perp=2\nu_0/\gamma_1$ introduced in
Eq.~\eqref{eq:length-X}, we define
\begin{equation}
X=
\frac{1}{\kappa\ell_\perp}
=
\frac{\gamma_1}{2\kappa\nu_0}.
\label{eq:zigzag-X}
\end{equation}
Evaluation of the integral then gives
\begin{equation}
\frac{w_l(\kappa)}{w_1(\kappa)}
=
\binom{2l-2}{l-1}X^{2l-2}.
\label{eq:zigzag-layer-weight}
\end{equation}
The dimensionless variable $X$ compares the amplitude penetration length
$\kappa^{-1}$ with the interlayer length $\ell_\perp$. Increasing the
penetration depth enhances the higher powers of $X$ and transfers normalized
weight toward later layers.

The connection with the lattice edge momentum follows from the exact
monolayer zigzag profile. A mode with dimensionless momentum $q$ along the
zigzag direction has the row-to-row decay factor
\begin{equation}
\lambda_q
=
-2\cos\left(\frac{q}{2}\right).
\label{eq:zigzag-lambda}
\end{equation}
Normalizability requires $|\lambda_q|<1$, corresponding to
$2\pi/3<q<4\pi/3$ modulo $2\pi$. If
$a_\perp=3a_{cc}/2$ is the separation between consecutive zigzag rows, the
exact amplitude decay constant associated with $q$ is
\begin{equation}
\kappa(q)
=
-\frac{1}{a_\perp}\ln|\lambda_q|.
\label{eq:zigzag-kappa-q}
\end{equation}
Near the projected valley endpoints, $|\lambda_q|\rightarrow1$ and
$\kappa(q)\rightarrow0$, so the boundary mode penetrates increasingly far
from the step. At $q=\pi$, $\lambda_q=0$, and the first-layer state is
restricted to the boundary row. 

\begin{figure*}[t]
\centering
\includegraphics[width=\textwidth]{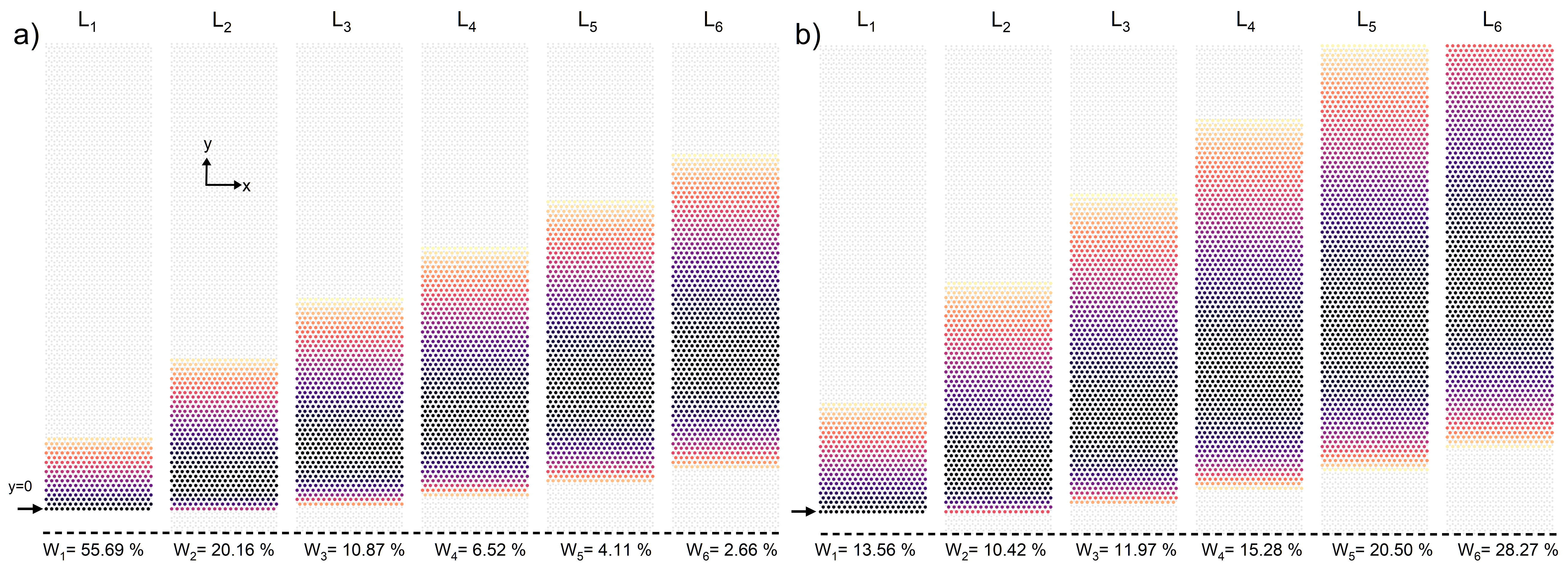}
\caption{\textbf{Momentum-controlled propagation of a zigzag boundary mode.}
Layer-resolved site density of the $B_1$-polarized zigzag zero mode in an
$N=6$ rhombohedral stack. The arrow marks the $L_1$ termination at $y=0$.
The probability density is uniform along the edge direction $x$, and only the $B_l$ sublattices carry zero-mode amplitude. \textbf{(a)}
$q=1.285\pi$, for which $\lambda_q=0.866$,
$\kappa^{-1}=14.77$~\AA, and $X=0.395$. The state remains concentrated near the initial surface, with $W_1=55.69\%$ and $W_6=2.66\%$.
\textbf{(b)} $q=1.300\pi$, for which $\lambda_q=0.91$,
$\kappa^{-1}=22.07$~\AA, and $X=0.59$. The increased penetration depth
transfers the largest layer weight to the remote surface, where
$W_6=28.27\%$. Each layer is independently color normalized to emphasize the spatial evolution of the mode; quantitative comparisons between layers are given by the integrated weights.}
\label{fig:zigzag-density}
\end{figure*}

Figure~\ref{fig:zigzag-density} illustrates the resulting redistribution
using the exact lattice profiles derived below. For $q=1.285\pi$, the mode has $X=0.395$ and remains predominantly on the initial surface, with
$W_1=55.69\%$ and $W_6=2.66\%$. At $q=1.300\pi$, the penetration depth
increases and $X=0.590$. The distribution then shifts across the stack:
$W_1$ decreases to $13.56\%$, while the remote layer acquires the largest
contribution, $W_6=28.27\%$. Simultaneously, the density maximum moves farther from the step in successive layers, displaying the combined polynomial growth and exponential decay predicted by Eq.~\eqref{eq:zigzag-propagated}.

\subsection{Exact lattice evaluation and selection of edge momenta}
\label{sec:zigzag-numerics}

The straight zigzag geometry allows an exact lattice treatment of the
boundary state. Translation symmetry along the edge makes the longitudinal momentum $q$ a conserved quantity, reducing the lattice zero-energy equations at each $q$ to a one-dimensional recursion between zigzag rows. The amplitudes on successive layers can therefore be obtained analytically, and their layer weights can be evaluated without truncating the transverse profile. This simplification relies on the straight, translationally invariant boundary: for the finite vacancy and void geometries considered above, the open boundary conditions couple different momenta and the complete lattice eigenstates are obtained numerically.

We index the zigzag rows by $n=0,1,\ldots$, starting at the termination and proceeding into the region containing $L_1$. The system is semi-infinite in the transverse direction and periodic along the edge. The layer weights are defined per zigzag unit cell and normalized using the complete transverse profile.  After removing the Bloch phase along the edge, the row amplitudes of the $B$-terminated mode satisfy
\begin{equation}
F_l(q,n)-\lambda_qF_l(q,n-1)=-\frac{\gamma_1}{\gamma_0}F_{l-1}(q,n),\qquad F_l(q,-1)=0,
\label{eq:zigzag-lattice-recursion}
\end{equation}
with $F_1(q,n)=\lambda_q^n$. Iteration gives
\begin{equation}
\psi_{B_l}(q,n)=\mathcal N_q\left(-\frac{\gamma_1}{\gamma_0}\right)^{l-1}\binom{n+l-1}{l-1}\lambda_q^n,\qquad n\geq0.
\label{eq:zigzag-amplitude-numerical}
\end{equation}
For large $n$, $\binom{n+l-1}{l-1}\simeq n^{l-1}/(l-1)!$. Together with $y=na_\perp$, $\nu_0=a_\perp\gamma_0$, and $|\lambda_q|^n=\exp[-\kappa(q)y]$, this reduces to Eq.~\eqref{eq:zigzag-propagated}, up to layer-dependent phases. The finite value of Eq.~\eqref{eq:zigzag-amplitude-numerical} at $n=0$ is a lattice-scale boundary contribution absent from the continuum approximation. Writing
\begin{equation}
r=\frac{\gamma_1}{\gamma_0},\qquad x_q=|\lambda_q|^2,
\label{eq:zigzag-r-x}
\end{equation}
the unnormalized weight on $L_l$ is
\begin{align}
w_l(q)&=r^{2(l-1)}\sum_{n=0}^{\infty}\binom{n+l-1}{l-1}^{\!2}x_q^n \nonumber\\
&=\frac{r^{2(l-1)}}{(1-x_q)^{2l-1}}\sum_{k=0}^{l-1}\binom{l-1}{k}^{\!2}x_q^k .
\label{eq:zigzag-exact-weight}
\end{align}
The second line is the closed expression used in the calculation, so no transverse cutoff is introduced. In the long-penetration regime, $\kappa a_\perp\ll1$, one has $1-x_q\simeq2\kappa a_\perp$. Equation~\eqref{eq:zigzag-exact-weight} then reduces to the continuum scaling in Eq.~\eqref{eq:zigzag-layer-weight}. The exact lattice expression is retained for the density maps and for every entry in Table~\ref{tab:zigzag-layer-weights}.

The calculations use the same parameters as before, with $N=6$. Because $|\lambda_{2\pi-q}|=|\lambda_q|$, the layer weights are symmetric about $q=\pi$. We therefore sample the nonredundant interval $\pi\leq q<4\pi/3$. The six tabulated momenta are equally spaced between $q=\pi$ and $q=4\pi/3-0.05$. The endpoint $4\pi/3$ is excluded because $|\lambda_q|\rightarrow1$, the penetration length diverges, and the semi-infinite transverse state is no longer normalizable. Table~\ref{tab:zigzag-layer-weights} shows the resulting redistribution. Between $q/\pi=1$ and $1.25$, $X$ increases from $0$ to $0.22$, while at least $85.82\%$ of the state remains on $L_1$. At $q/\pi=1.31$, the penetration length reaches $47.7$~\AA\ and $X=1.27$. The state then becomes strongly polarized toward the opposite surface: $83.6\%$ of the weight resides on $L_6$ and $13.6\%$ on $L_5$. This crossover is produced by the increasing penetration depth of the first-layer boundary mode. At $q=\pi$, the first-layer seed is restricted to $n=0$, but the exact lattice recursion produces amplitudes on the rows aligned with the step in subsequent layers. The small nonzero weights on $L_2$ and $L_3$ are short-distance lattice contributions.

\begin{table*}[t]
\caption{\textbf{Layer-resolved weight of the zigzag zero mode.} Exact lattice results for a $B_1$-polarized zigzag mode in an $N=6$ rhombohedral stack. The edge momentum $q$ determines the decay factor $\lambda_q$, the penetration length $\kappa^{-1}$, and the dimensionless propagation scale $X$. The remaining columns give the corresponding normalized weights.}
\label{tab:zigzag-layer-weights}
\centering
\begin{ruledtabular}
\begin{tabular}{cccccccccc}
$q/\pi$ & $\lambda_q$ & $\kappa^{-1}$ (\AA) & $X$ & $W_1$ (\%) & $W_2$ (\%) & $W_3$ (\%) & $W_4$ (\%) & $W_5$ (\%) & $W_6$ (\%) \\
\hline
1.000 & 0     & 0      & 0     & 98.702 & 1.281  & 0.017 & 0 & 0     & 0     \\
1.063 & 0.199 & 1.320  & 0.035 & 98.535 & 1.442  & 0.023 & 0 & 0     & 0     \\
1.127 & 0.396 & 2.301  & 0.062 & 97.876 & 2.068  & 0.054 & 0.002 & 0 & 0     \\
1.190 & 0.589 & 4.030  & 0.108 & 95.826 & 3.935  & 0.223 & 0.014 & 0.001 & 0 \\
1.254 & 0.777 & 8.432  & 0.225 & 85.827 & 11.353 & 2.207 & 0.478 & 0.109 & 0.026 \\
1.317 & 0.956 & 47.770 & 1.277 & 0.023  & 0.078  & 0.401 & 2.280 & 13.612 & 83.606 \\
\end{tabular}
\end{ruledtabular}
\end{table*}

For the density maps in Fig.~\ref{fig:zigzag-density}, ten unit cells are repeated along the zigzag direction, while the transverse rows define only the displayed window. The normalization and layer weights use the complete analytical sums in Eq.~\eqref{eq:zigzag-exact-weight} and are independent of this plotting window. Each layer is displayed on a separate logarithmic scale.

\section{Moir\'e and periodic interface systems}
\label{sec:interfaces}
The vacancy and boundary examples show how a state generated in $L_1$
can acquire weight on successive layers. We now consider the same
connection in periodic systems: an ABC stack subject to a pseudogauge
field on $L_1$, and an ABC stack coupled to a twisted monolayer. In each
case, the interface determines the electronic states and their momentum
harmonics, while the ABC stack determines how their amplitudes are
related across the layers. We first derive the common momentum-dependent propagation scale and the surface projection. We then specify the two interface Hamiltonians, their numerical layer weights, and their real-space textures. The exact layer recursion applies to sublattice-polarized zero-energy states. For finite-energy bands, we calculate the layer weights from the full Hamiltonian and use the zero-energy propagation scale to interpret the results. 

\subsection{Common framework}
\label{sec:periodic-common}

\subsubsection{Surface projection and layer reconstruction}
\label{sec:projection}

In the minimal model of a rhombohedral multilayer, the low-energy surface basis is given by $\mathcal L=\{A_1,B_N\}$, while the remaining sites form the dimerized part
\begin{equation}
\mathcal H=\{B_1,A_2,B_2,A_3,\ldots,B_{N-1},A_N\}.
\label{eq:sectors}
\end{equation}
The exact energy-dependent Schur complement is
\begin{equation}
H_{\mathrm{eff}}(E)=H_{LL}+H_{LH}(E-H_{HH})^{-1}H_{HL}.
\label{eq:Schur}
\end{equation}
The leading zero-energy expression can be obtained directly from the unique path through the dimers. We define
\begin{equation}
\begin{aligned}
D_{N-1}(\bm p)&=
\left[-\frac{\nu_0\pi(\bm p)}{\gamma_1}\right]^{N-1},\\
\mathcal X_N(\bm p)&=\nu_0\pi(\bm p)
D_{N-1}(\bm p)
=(-1)^{N-1}\frac{[\nu_0\pi(\bm p)]^N}
{\gamma_1^{N-1}},
\end{aligned}
\label{eq:DX}
\end{equation}
where the factor $\pi^N$ carries momentum-space winding $N$. The low-energy Hamiltonian in the $\{A_1,B_N\}$ basis and the corresponding reconstruction of $B_1$ are given by
\begin{equation}
H_{\mathrm{surf}}^{(N)}(\bm p)=
\begin{pmatrix}
0&\mathcal X_N^\dagger\\
\mathcal X_N&0
\end{pmatrix},
\qquad
\psi_{B_1}=D_{N-1}^\dagger\psi_{B_N}\textcolor{red}{,}
\label{eq:Hsurface}
\end{equation}
where the real-space operator in Eq.~\eqref{eq:P-operator} becomes
\begin{equation}
\mathcal P(\bm p)=-\frac{\gamma_1}
{\nu_0\pi^\dagger(\bm p)},
\qquad
[D_{N-1}(\bm p)]^\dagger
=\mathcal P^{-(N-1)}(\bm p).
\label{eq:P-D-relation}
\end{equation}
Thus $D_{N-1}^{\dagger}$ reconstructs the $B_1$ component from $B_N$,
whereas $\mathcal P^{N-1}$ continues $B_1$ toward $B_N$. As shown below, in both systems, both the surface kinetic term and the projected interface term contain the same overall power $\gamma_1^{1-N}$.  Changing $\gamma_1$ therefore rescales that projected two-surface Hamiltonian without, by itself, changing its surface-basis eigenvectors.  The physical layer weights nevertheless depend on $\gamma_1$ because the eliminated dimer amplitudes must be reconstructed.  This distinction is why the complete eigenvectors, rather than $D_{N-1}$ alone, are used in the numerical layer-weight analysis below.

For a general interface vertex $v_{ij}$ that converts incoming harmonic $j$
to outgoing harmonic $i$, its projection onto the basis $(A_1,B_N)$ is
\begin{equation}
v_{ij}^{\mathrm{surf}}=
\begin{pmatrix}
v_{ij}^{AA}&v_{ij}^{AB}D_j^\dagger\\
D_i v_{ij}^{BA}&D_i v_{ij}^{BB}D_j^\dagger
\end{pmatrix},
\qquad D_i=D_{N-1}(\bm p_i),
\label{eq:general-vertex}
\end{equation}
which makes the momentum carried by each interface component explicit. The interface determines the state and its harmonic amplitudes. For an
exact zero mode with a nonzero $B_1$ component, the layer equations fix
the continuation of that component through the unmodified slab. A state
without $B_1$ overlap does not enter this channel.

\subsubsection{Momentum-dependent propagation}
\label{sec:beta}

For each nonzero momentum channel, define
\begin{equation}
\beta(\bm p)=|\mathcal P(\bm p)|
=\frac{\gamma_1}{\nu_0|\bm p|},
\qquad
\frac{|\psi_{B_N}(\bm p)|}{|\psi_{B_1}(\bm p)|}
=\beta^{N-1}(\bm p).
\label{eq:beta}
\end{equation}
For $\beta(\bm p)<1$, the amplitude of the selected harmonic decreases
from $B_1$ toward $B_N$; for $\beta(\bm p)>1$, it increases. The power
$N-1$ makes this momentum selectivity more pronounced across thicker
stacks. For example, in the twisted system, the conventional parameter
$\beta_0=\gamma_1/(\nu_0k_\theta)$ uses the Dirac-point separation
$k_\theta$, defined in Eq.~\eqref{eq:alpha-twist} below. Each harmonic contributes
\begin{equation}
\beta(\bm p_i)=\beta_0\frac{k_\theta}{|\bm p_i|}.
\label{eq:beta-harmonic}
\end{equation}
The layer distribution of a Bloch state is therefore governed by the momentum-selective propagation of its individual harmonics. Each harmonic samples a different $\beta(\bm p_i)$ and is redistributed across the stack according to its momentum. Their combined amplitudes determine whether the state remains near the interface or acquires substantial weight on the remote surface. In consequence, a single parameter $\beta_0$ cannot determine the layer polarization of the complete state.

In the pseudogauge and twisted systems the multilayer momenta are, respectively, $\bm p_i^{\mathrm g}=\bm k+\bm G_i$ and $\bm p_i^{\mathrm{tw}}=\bm k-\bm K_R+\bm G_i$. The interface couples these harmonics, and their amplitudes must be determined from the full Hamiltonian. 
 We write
\begin{equation}
\bm G_{mn}=m\bm G_1+n\bm G_2,
\qquad
\bm k-\bm K_R^s=c_1^s\bm G_1+c_2^s\bm G_2.
\label{eq:momentum-coefficients}
\end{equation}
For reciprocal primitive vectors separated by $120^\circ$,
\begin{equation}
|\bm p_{mn}^s|^2=G_s^2
[(m+c_1^s)^2+(n+c_2^s)^2
-(m+c_1^s)(n+c_2^s)]
\equiv G_s^2\,\mathcal Q_{mn}^s.
\label{eq:p-magnitude}
\end{equation}
The channel condition $\beta=1$ is therefore
\begin{equation}
L_s^*=\frac{4\pi\nu_0}{\sqrt{3}\gamma_1}
\sqrt{\mathcal Q_{mn}^s}
=\frac{4\pi\nu_0}{\sqrt{3}\gamma_1}
\sqrt{(m+c_1^s)^2+(n+c_2^s)^2
-(m+c_1^s)(n+c_2^s)},
\label{eq:beta-boundary}
\end{equation}
which is the expression plotted in Fig.~2 of the main text.
Here $L_s=L_{\mathrm M}$ and $\bm K_R^{\mathrm g}=0$ for the gauge model,
whereas $L_s=L_M$ and $\bm K_R^{\mathrm{tw}}=\bm K$ for the twisted model.
Three representative first-star vectors are indexed by $(m,n)=(1,0),(0,1),(-1,-1)$. Equation~\eqref{eq:beta-boundary} gives the condition
$\beta(\bm p)=1$ for a single Fourier harmonic. It therefore implies $|\psi_{B_N}(\bm p)|=|\psi_{B_1}(\bm p)|$ for that harmonic only. A Bloch eigenstate contains several momentum harmonics $\bm p_i^s$, each with its own propagation factor $\beta(\bm p_i^s)$, so this condition does
not imply equal total weights on $L_1$ and $L_N$. When $\bm p_i=0$, Eq.~\eqref{eq:beta} cannot be applied because it was obtained by inverting the Dirac operator, which vanishes at that momentum. The corresponding component must instead be determined directly from the original zero-energy equations. In the chiral twisted model, the apparent pole is cancelled by the Bloch-state prefactor, leaving a finite wavefunction.

The moir\'e length provides a direct way to control propagation through the stack. At fixed reduced Bloch momentum and harmonic indices, increasing the moir\'e length reduces $|\bm p|$ and therefore increases $\beta(\bm p)$. Long-period moir\'e modulations can thus enhance the relative amplitude of an interface harmonic on the remote surface. Equation~\eqref{eq:beta-boundary} identifies the moir\'e length at which that harmonic reaches $\beta=1$, separating the regimes where its amplitude decreases or increases across the stack. The crossover occurs at different moir\'e lengths for different harmonics and Bloch momenta. The layer polarization of the complete state consequently depends on how these harmonics are populated, which also changes with the interface Hamiltonian as the moir\'e length is varied.

\subsection{Periodic transverse pseudogauge field}
\label{sec:gauge}
\subsubsection{Interface potential and zero-mode structure}

Let $\bm g_1,\bm g_2$, and $\bm g_3=-(\bm g_1+\bm g_2)$ be the positive
first-star vectors of a triangular superlattice with moir\'e period
$L_{\mathrm M}$, where
\begin{equation}
|\bm g_j|=G=\frac{4\pi}{\sqrt{3}L_{\mathrm M}}.
\label{eq:gauge-period}
\end{equation}
Define 
\begin{equation}
f_1(\bm r)=2\sum_{j=1}^{3}\cos(\bm g_j\cdot\bm r),
\qquad
f_2(\bm r)=-2\sum_{j=1}^{3}\sin(\bm g_j\cdot\bm r).
\label{eq:f-functions}
\end{equation}
A generic transverse pseudogauge potential is given by
\begin{equation}
V_{\mathrm g}(\bm r)=
\left[\widehat{\bm z}\times
\frac{u_\perp\nabla f_2(\bm r)
+\widetilde u_\perp\nabla f_1(\bm r)}{G}\right]\cdot\bm\sigma,
\label{eq:Vg}
\end{equation}
where $\bm\sigma=(\sigma_x,\sigma_y)$. We set $\widetilde u_\perp=0$ to consider only a simple even triangular pseudogauge field, although our results are also valid for a general field. This field acts only on $L_1$. Such pseudogauge fields arise from nonuniform strain
and  relaxation~\cite{Vozmediano2010Gauge,Guinea2010Strain,SanJose2014Electronic}.
For $\bm q_{j,s}=s\bm g_j$, $s=\pm1$, and
$\bm e_{j,s}=\bm q_{j,s}/G$, Eq.~\eqref{eq:Vg} has the Fourier representation
\begin{equation}
V_{\mathrm g}(\bm r)=
\sum_{j=1}^{3}\sum_{s=\pm1}V_{j,s}
\mathrm{e}^{\mathrm{i}\bm q_{j,s}\cdot\bm r},
\label{eq:Vg-Fourier}
\end{equation}
where
\begin{equation}
V_{j,s}=
\mathrm{i}(\widetilde u_\perp+\mathrm{i}s u_\perp)
(e_{j,s,x}\sigma_y-e_{j,s,y}\sigma_x),
\qquad
V_{j,-}=V_{j,+}^\dagger.
\label{eq:Vg-matrices}
\end{equation}
The full Hamiltonian is given by
\begin{equation}
H^{\mathrm g}=H_N^{\mathrm{ABC}}
+|L_1\rangle\langle L_1|\otimes V_{\mathrm g}(\bm r)
+H_m^{(\mathrm g)}.
\label{eq:Hg}
\end{equation}
A smooth periodic zero-flux Abelian vector potential has exact zero modes at the folded Dirac point but does not generically pin a zero-energy eigenvalue throughout the Brillouin zone.  Strong fields can nevertheless make the central bands very narrow, consistent with related periodically strained graphene models~\cite{Gao2023Strain,Wan2023Nearly}. Note that for the numerical spectra of both periodic systems, we introduce a new term $H_m$ which denotes a uniform sublattice mass $m_0\sigma_z$ on every layer. At an exact zero mode, it shifts the two sublattice sectors without changing their wavefunctions or layer weights and is introduced only to visualize the effect of the charge transfer. 

\begin{figure}
\centering
\includegraphics[width=0.96\textwidth]{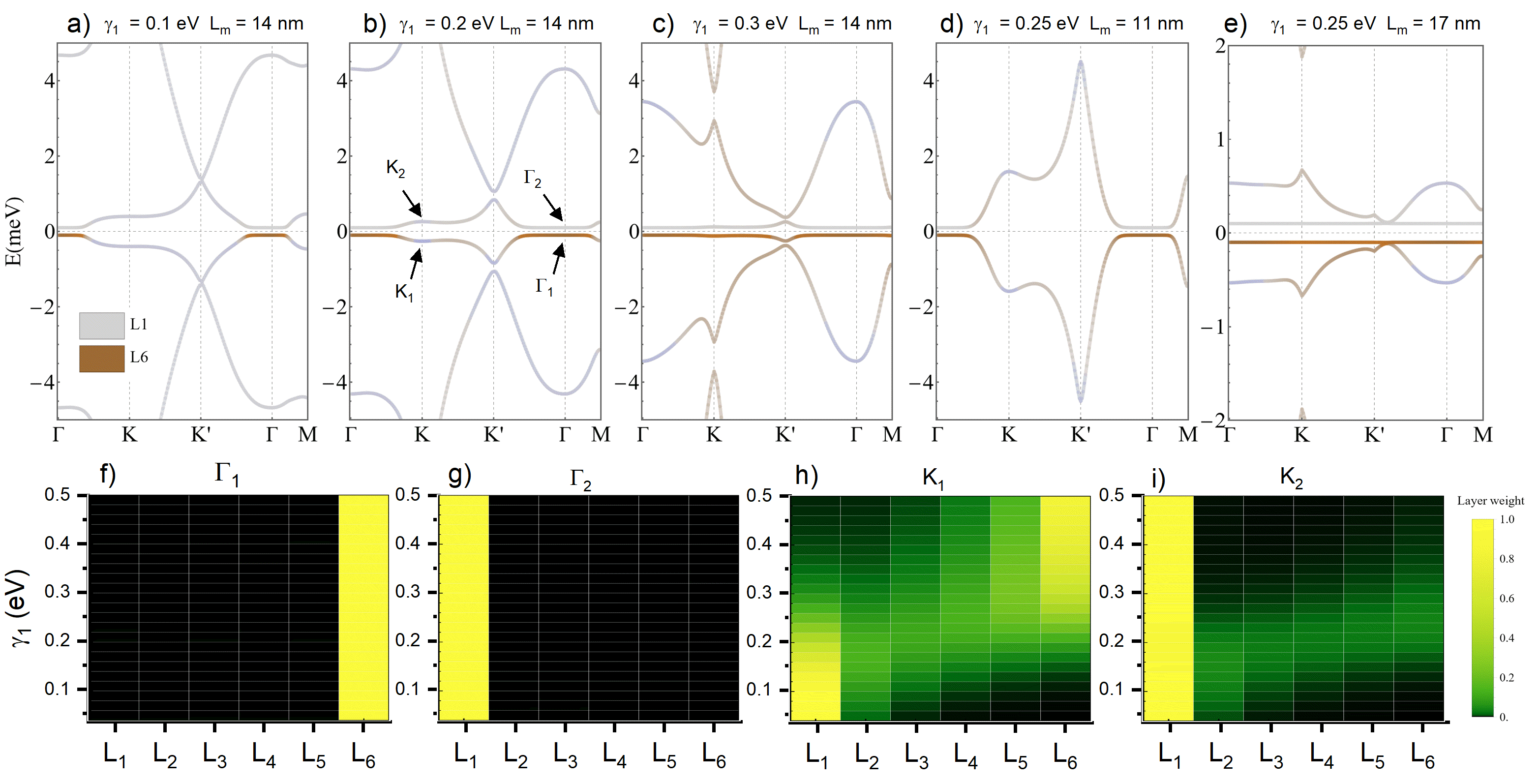}
\caption{\textbf{Layer redistribution under a periodic pseudogauge field.}
Band structures for $N=6$, $|m_0|=0.1$ meV, $u_\perp=300$ meV, and
$\widetilde u_\perp=0$.  Panels (a)--(c) use $L_{\mathrm M}=14$ nm and
$\gamma_1=0.1$, $0.2$, and $0.3$ eV; panels (d) and (e) use
$\gamma_1=0.25$ eV and $L_{\mathrm M}=11$ and $17$ nm.  Line colors blend the normalized layer weights from gray ($L_1$) to brown ($L_6$).  At $L_{\mathrm M}=14$ nm, panels (f)--(i) show the layer weights of the lower and upper central states $\Gamma_{1,2}$ and $K_{1,2}$ versus $\gamma_1$. The lower state redistributes from $L_1$ toward $L_6$ at $K$, whereas the upper state remains interface biased.  At $\Gamma$, the two central states retain opposite-surface character.}
\label{fig:BandsGauge}
\end{figure}

\subsubsection{Plane-wave and projected Hamiltonians}
The pseudogauge momenta are
\begin{equation}
\bm p_i^{\mathrm g}=\bm k+\bm G_i,
\label{eq:pg}
\end{equation}
and the matrix elements are
\begin{align}
[H^{\mathrm g}(\bm k)]_{il,jl'}={}&
\delta_{ij}[H_N^{\mathrm{ABC}}(\bm p_i^{\mathrm g})
+H_m^{(\mathrm g)}]_{ll'}\nonumber\\
&+\delta_{l1}\delta_{l'1}
\sum_{a=1}^{3}\sum_{s=\pm1}
\delta_{\bm G_i-\bm G_j,\bm q_{a,s}}V_{a,s}.
\label{eq:Hg-plane-wave}
\end{align}
For the pseudogauge system, define
$g_{ij}=\langle\bm G_i,A_1|V_{\mathrm g}|
\bm G_j,B_1\rangle$ and
$\mathcal X_i=\mathcal X_N(\bm p_i^{\mathrm g})$. The projected block is
\begin{equation}
\begin{aligned}
[H_{\mathrm{proj}}^{\mathrm g}]_{ij}={}&
\delta_{ij}
\begin{pmatrix}0&\mathcal X_i^\dagger\\\mathcal X_i&0\end{pmatrix}
+
\begin{pmatrix}
0&g_{ij}D_j^\dagger\\
[g_{ji}D_i^\dagger]^*&0
\end{pmatrix}\textcolor{red}{,}
\end{aligned}
\label{eq:Hg-projected}
\end{equation}
where $D_j$ is given in Eq.~\eqref{eq:DX}. 
\subsubsection{Numerical spectra and layer redistribution}
\label{sec:gauge-numerics}

Figure~\ref{fig:BandsGauge} shows the pseudogauge spectra and the layer
weights obtained from the full plane-wave eigenvectors. At fixed
$L_{\mathrm M}=14$~nm, panels (a)--(c) show the evolution as $\gamma_1$ increases. Panels (d) and (e) vary $L_{\mathrm M}$ at fixed $\gamma_1$. Both changes clearly modify the layer distribution, indicated by color, as expected from the change in the channel-dependent $\beta$ factors. 

The layer weights in panels (f)--(i) reveal the momentum dependence.
At $\Gamma$, the two central states $\Gamma_1$ and $\Gamma_2$ remain
polarized on $L_6$ and $L_1$, respectively. At $K$, the lower state $K_1$ evolves from $L_1$ through the intermediate layers toward $L_6$, whereas $K_2$ remains predominantly on the interface side. The real-space densities in
Fig.~\ref{fig:Density}(a),(b) complement these layer weights.

\subsection{Chiral twisted monolayer--ABC interface}
\label{sec:twist}
\subsubsection{Geometry, Hamiltonian, and exact flat bands}

We now consider a twisted monolayer--rhombohedral graphene system. The monolayer $L_0$ is rotated by $+\theta/2$, while the ABC slab is rotated
by $-\theta/2$.  The reciprocal moiré vectors are given by
\begin{equation}
\bm G_i=R(\theta/2)\bm b_i-R(-\theta/2)\bm b_i,
\qquad
L_M=\frac{a}{2\sin(\theta/2)},
\label{eq:twisted-geometry}
\end{equation}
where $\bm b_i$ are the graphene reciprocal vectors. Choose $\bm Q_1=0$, $\bm Q_2=\bm G_1$, and $\bm Q_3=\bm G_1+\bm G_2$.  The interlayer tunneling is
\begin{equation}
\begin{aligned}
T_{01}(\bm r)&=
\sum_{j=1}^{3}T_j
\mathrm{e}^{-\mathrm{i}\bm Q_j\cdot\bm r},\\
T_j&=u'
\begin{pmatrix}
0&\mathrm{e}^{-\mathrm{i}\phi_j}\\
\mathrm{e}^{\mathrm{i}\phi_j}&0
\end{pmatrix},
\qquad
\phi_j=(j-1)\frac{2\pi}{3}.
\end{aligned}
\label{eq:T01}
\end{equation}
This is the usual continuum representation~\cite{Lopes2007Graphene,Bistritzer2011Moire,Koshino2015Interlayer}, in the chiral approximation.  At the magic angle this
model supports exact zero-energy flat bands~\cite{Tarnopolsky2019_origin,Ledwith2022Family,Wang2022Hierarchy}. The  $(N+1)$-layer Hamiltonian is
\begin{equation}
H^{\mathrm{tw}}=
\begin{pmatrix}
h_0&T_{01}&0&\cdots&0\\
T_{01}^\dagger&h_1&T_{\gamma_1}&\ddots&\vdots\\
0&T_{\gamma_1}^\dagger&h_2&\ddots&0\\
\vdots&\ddots&\ddots&\ddots&T_{\gamma_1}\\
0&\cdots&0&T_{\gamma_1}^\dagger&h_N
\end{pmatrix},
\label{eq:Htw}
\end{equation}
For the numerical electronic structure, the uniform sublattice mass introduced above is added to $H^{\mathrm{tw}}$. Exact flatness creates the zero-energy subspace; continuation across the ABC slab additionally requires finite overlap with the $B_1$-connected sector. To make this explicit, we introduce
\begin{equation}
k_\theta=2K_D\sin\frac{\theta}{2},
\qquad E_\theta=\nu_0k_\theta,
\qquad \alpha=\frac{u'}{E_\theta},
\qquad K_D=\frac{4\pi}{3a}.
\label{eq:alpha-twist}
\end{equation}
After ordering the sites by sublattice, the strict chiral Hamiltonian takes
the form $H_{\mathrm{ch}}=\bigl(\begin{smallmatrix}0&\mathcal D_N^\dagger\\
\mathcal D_N&0\end{smallmatrix}\bigr)$.  At a generic Bloch momentum its
zero-mode block can be arranged as
\begin{equation}
\mathcal D_N=
\begin{pmatrix}
\mathcal D_{\mathrm T}&C\\
0&R_{N-1}
\end{pmatrix},
\label{eq:chiral-block-triangular}
\end{equation}
where $\mathcal D_{\mathrm T}$ is precisely the chiral twisted-bilayer
operator formed by $L_0$ and $L_1$, and $R_{N-1}$ is the remaining triangular
ABC chain.  At generic momentum, $R_{N-1}$ is invertible, so the zero-mode problem
reduces to that of $\mathcal D_{\mathrm T}$. The construction below extends
these modes to the full slab. The exact flat bands therefore occur at the chiral twisted-bilayer magic angle
$\alpha=\alpha_j$, independently of $N$ and $\gamma_1$; the first value in the
convention of Eq.~\eqref{eq:alpha-twist} is $\alpha_1\simeq0.586$~\cite{Tarnopolsky2019_origin}.
The $A$-polarized mode is the chiral twisted-bilayer state embedded on $L_0$ and $L_1$, with zero amplitude on $A_2,\ldots,A_N$.  The $B$-polarized mode begins with the other chiral twisted-bilayer zero mode and satisfies the ABC recursion. For the harmonic
$\bm p_{\bm G}=\bm k-\bm K_R+\bm G$, its slab components may be written as
\begin{equation}
u_{B_l,\bm k}(\bm G)=\lambda_{N,\bm k}
\left[-\frac{\gamma_1}
{\nu_0\pi^\dagger(\bm p_{\bm G})}\right]^{l-1}
u_{B_1,\bm k}^{\mathrm T}(\bm G),
\qquad l=1,\ldots,N.
\label{eq:exact-propagated-mode}
\end{equation}
The inverse Dirac factors appear singular at the folded slab Dirac point.
However, the Bloch prefactor $\lambda_{N,\bm k}$ has a zero of order $N-1$ at
that point, which cancels the highest-order pole and makes every component of
the exact state regular~\cite{Ledwith2022Family}.  Thus $\gamma_1$ controls the layer distribution of the propagated flat-band state, but not the magic condition that creates the two flat bands.

\begin{figure}[t]
\centering
\includegraphics[width=0.65\textwidth]{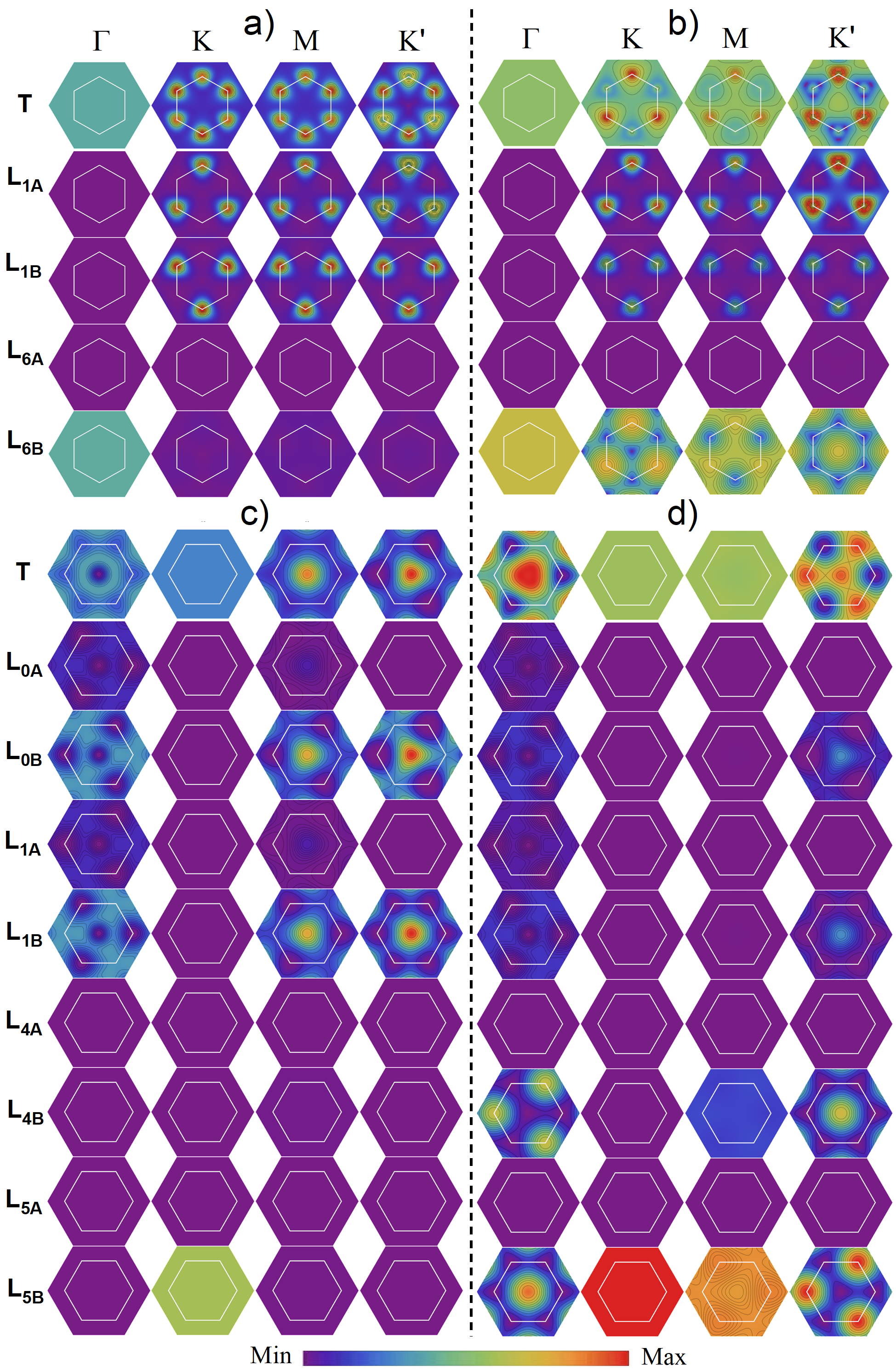}
\caption{\textbf{Real-space structure of the propagated states.}
Layer- and sublattice-resolved charge density of the lower middle band for selected layers. For the periodic pseudogauge system, panels (a) and (b) show the density before and after the propagation crossover, respectively. Panels (c) and (d) show the corresponding densities for the twisted system. The columns correspond to $\Gamma$, $K$, $M$, and $K'$. In each panel, the row labeled T is the probability density summed over all layers and sublattices. The pseudogauge system contains $N=6$ layers. The twisted system contains the rotated monolayer $L_0$ and an $N=5$ ABC slab. The remote component is concentrated on the undimerized $B_N$ site, with negligible weight on $A_N$. In the pseudogauge panels, the $B_N$ density at $\Gamma$ is nearly uniform.}
\label{fig:Density}
\end{figure}

\subsubsection{Plane-wave and projected Hamiltonians}
For the twisted system, the monolayer and slab momenta are
\begin{equation}
\begin{aligned}
\bm p_i^0&=R(\theta/2)(\bm k+\bm G_i-\bm K_0),\\
\bm p_i^R&=R(-\theta/2)(\bm k+\bm G_i-\bm K_R).
\end{aligned}
\label{eq:twisted-momenta}
\end{equation}
Up to the common rotation, $\bm p_i^R$ is the momentum set
$\bm p_i^{\mathrm{tw}}=\bm k-\bm K_R+\bm G_i$ quoted in the main text.
The tunneling matrix elements are 
\begin{equation}
\langle\bm G_i,0|T_{01}|\bm G_j,1\rangle
=\sum_{m=1}^{3}T_m
\delta_{\bm G_i-\bm G_j,-\bm Q_m}.
\label{eq:T01-selection}
\end{equation}
The surface projection makes the effect of the twisted interface on
the remote surface explicit. Eliminating the dimerized orbitals of the
ABC slab while retaining the twisted monolayer gives, in the basis
$(A_0,B_0,A_1,B_N)$,
\begin{equation}
H_{\mathrm{proj}}^{\mathrm{tw}}=
\begin{pmatrix}
0&\nu_0\pi_0^\dagger&0&T_{A_0B_1}D^\dagger\\
\nu_0\pi_0&0&T_{B_0A_1}&0\\
0&T_{A_1B_0}&0&\mathcal X_N^\dagger\\
DT_{B_1A_0}&0&\mathcal X_N&0
\end{pmatrix}.
\label{eq:Htw-projected}
\end{equation}
Here $D\equiv D_{N-1}$ acts on the slab momentum harmonics, and
$\mathcal X_N$ describes the coupling between the two outer
sublattices of the ABC slab. The twisted monolayer couples directly
to $A_1$ through $T_{B_0A_1}$, while its coupling to the remote
surface $B_N$ arises from the interface hopping to $B_1$ and the
factor $D$. Thus, the projected Hamiltonian shows how a coupling
localized at the twisted interface is connected to the opposite
surface through the chiral structure of the stack. This connection
is momentum dependent: for each harmonic, $D$ is related to the
layer-transfer scale by $|D|=\beta^{-(N-1)}$.

\subsubsection{Numerical spectra and layer redistribution}
\label{sec:twist-numerics}
Figure~3 of the main text shows the twisted $N+1$ system with an $N=5$ ABC
slab, $\gamma_0=3.16$ eV, $u'=0.0975$ eV, and $|m_0|=0.05$ meV.
At $L_M=16.9$ nm, the spectra use $\gamma_1=0.05$, $0.15$, and
$0.20$ eV; the period sweep uses $\gamma_1=0.185$ eV and $L_M=14$, $16$,
and $18$ nm. The Hamiltonian is rediagonalized at each parameter value and
the layer weights are calculated by defining
\begin{equation}
W_l=\sum_{\alpha=A,B}\sum_{\bm G}|u_{l\alpha}(\bm G)|^2,
\qquad \sum_l W_l=1\textcolor{red}{.}
\label{eq:bloch-layer-norm}
\end{equation}
Increasing $\gamma_1$ or $L_M$ enhances the layer-step magnitude of a
specified nonzero slab harmonic. The lower central state develops weight
on the remote $L_5$ surface, while the upper state remains predominantly
interface polarized in the displayed finite-mass spectra. The weights at
$\Gamma$, $K$, and $K'$ resolve the momentum dependence and the intermediate
layer contributions near the crossover. This is a continuous change in layer
polarization within the energy-ordered central bands, not a band inversion.

The corresponding real-space densities are the twisted-system panels
(d),(e) of Fig.~\ref{fig:Density}. They resolve how probability transfers
from $L_0,L_1$ to the remote $B_N$ component and whether that component is
uniform or modulated within the moir\'e cell, as discussed in
Sec.~\ref{sec:numerics}.

\subsection{Comparison of real-space textures}
\label{sec:numerics}
The layer weights establish where the state resides across the stack, but do not show how its probability is distributed within the superlattice unit cell. We show this effect in Fig.~\ref{fig:Density} for the lower middle band in real space, by layer and sublattice, for representative parameters on the two sides of the propagation crossover. The maps resolve both the layer and sublattice character of the states. Note that weight transferred toward the remote surface is concentrated on $B_N$, while $A_N$ remains weak. It is also evident that the spatial texture depends on momentum: the remote weight can be nearly uniform at one high-symmetry point and strongly modulated at another. The interface Hamiltonian fixes the harmonic amplitudes, and their
coherent superposition determines the real-space density.

\section{Trigonal warping}
\label{sec:warping}

\subsection{Surface wavefunctions and notation}
\label{sec:warping-states}

Here we consider the effect of interlayer hoppings that break the isotropy of the model and induce trigonal warping, usually denoted by $\gamma_3$ and $\gamma_4$. These hoppings lead to momentum-dependent terms in the Hamiltonian and vanish as $|\bm k|\rightarrow0$. As mentioned above, the perturbations projected from one surface to the other have low momenta, so we expect these interlayer terms to have a weak effect on the phenomena discussed previously.

We obtain a more quantitative analysis by projecting these hoppings onto the two low-energy wavefunctions, which are mostly localized at the two surfaces of the stack and decay exponentially away from them. The two modes at zero energy obtained in the absence of $\gamma_3$ and $\gamma_4$ are
\begin{align}
|\psi_{A,\bm k}\rangle
&=\frac{A_{\bm k}}{|A_{\bm k}|}\mathcal C_{\bm k}^{-1}
\sum_{n=0}^{N-1}\rho_{\bm k}^{n}|L_{n+1},A,\bm k\rangle,
\nonumber\\
|\psi_{B,\bm k}\rangle
&=\mathcal C_{\bm k}^{-1}
\sum_{n=0}^{N-1}\rho_{\bm k}^{N-n}|L_{n+1},B,\bm k\rangle,
\label{eq:warping-states}
\end{align}
where $A_{\bm k}$ is the amplitude of the function in the first layer and $\mathcal C_{\bm k}$ is a normalization constant, with
\begin{equation}
\rho_{\bm k}=-\frac{\gamma_1}
{\nu_0 k\mathrm e^{\mathrm i\varphi_{\bm k}}},
\qquad
\mathcal C_{\bm k}=\sqrt{\frac{1-|\rho_{\bm k}|^{2N}}{1-|\rho_{\bm k}|^2}}.
\label{eq:warping-definitions}
\end{equation}
Here $\bm k$ is measured from the chosen Dirac point, $k=|\bm k|$, $\varphi_{\bm k}=\arg(k_x+\mathrm i k_y)$, and $n=0,\ldots,N-1$ corresponds to layers $L_{n+1}$. The layer factor $\rho_{\bm k}$ and normalization factor $\mathcal C_{\bm k}$ are the quantities denoted by $\beta$ and $C$ in the surface-state construction. These symbols distinguish them from the twisted-interface coupling $\alpha=u'/E_\theta$ and the vacancy amplitude $C$ used above.

\subsection{Projection of $\gamma_3$}

The term defined by $\gamma_3$ describes hoppings between neighboring layers and different sublattices, so it hybridizes the two surface-localized states, which reside on different sublattices. We obtain
\begin{align}
t_3(\bm k)
&\equiv\langle\psi_{A,\bm k}|H_{\gamma_3,\bm k}|\psi_{B,\bm k}\rangle
\nonumber\\
&=\frac{A_{\bm k}}{|A_{\bm k}|}\mathcal C_{\bm k}^{-2}\nu_3 k
\mathrm e^{\mathrm i\varphi_{\bm k}}
\sum_{n=0}^{N-2}(\rho_{\bm k}^*)^n\rho_{\bm k}^{N-n-2}
\nonumber\\
&=\frac{A_{\bm k}}{|A_{\bm k}|}\mathcal C_{\bm k}^{-2}\nu_3 k
\mathrm e^{\mathrm i\varphi_{\bm k}}\rho_{\bm k}^{N-2}
\frac{1-\mathrm e^{2\mathrm i(N-1)\varphi_{\bm k}}}
{1-\mathrm e^{2\mathrm i\varphi_{\bm k}}},
\qquad \nu_3=\frac{\sqrt{3}}{2}a\gamma_3.
\label{eq:warping-gamma3}
\end{align}
The intermediate sum follows from
\begin{equation}
\frac{\rho_{\bm k}^*}{\rho_{\bm k}}=\mathrm e^{2\mathrm i\varphi_{\bm k}},
\qquad
\sum_{n=0}^{N-2}(\rho_{\bm k}^*)^n\rho_{\bm k}^{N-n-2}
=\rho_{\bm k}^{N-2}\sum_{n=0}^{N-2}\mathrm e^{2\mathrm i n\varphi_{\bm k}}.
\label{eq:warping-angular-sum}
\end{equation}
Thus the angular quotient in Eq.~\eqref{eq:warping-gamma3} is a finite geometric sum. At $\mathrm e^{2\mathrm i\varphi_{\bm k}}=1$, its continuous value is $N-1$.

\subsection{Projection of $\gamma_4$}

The term proportional to $\gamma_4$, which does not hybridize the sublattices, is
\begin{align}
\varepsilon_4(\bm k)
&\equiv\langle\psi_{A,\bm k}|H_{\gamma_4,\bm k}|\psi_{A,\bm k}\rangle
=\langle\psi_{B,\bm k}|H_{\gamma_4,\bm k}|\psi_{B,\bm k}\rangle
\nonumber\\
&=\mathcal C_{\bm k}^{-2}\nu_4 k\mathrm e^{\mathrm i\varphi_{\bm k}}
\sum_{n=0}^{N-2}(\rho_{\bm k}^*)^n\rho_{\bm k}^{n+1}
\nonumber\\
&=\mathcal C_{\bm k}^{-2}\nu_4 k\mathrm e^{\mathrm i\varphi_{\bm k}}
\rho_{\bm k}\frac{1-|\rho_{\bm k}|^{2(N-1)}}{1-|\rho_{\bm k}|^2},
\qquad \nu_4=\frac{\sqrt{3}}{2}a\gamma_4.
\label{eq:warping-gamma4}
\end{align}
The intermediate sum is
\begin{equation}
\sum_{n=0}^{N-2}(\rho_{\bm k}^*)^n\rho_{\bm k}^{n+1}
=\rho_{\bm k}\sum_{n=0}^{N-2}|\rho_{\bm k}|^{2n}
=\rho_{\bm k}\frac{1-|\rho_{\bm k}|^{2(N-1)}}{1-|\rho_{\bm k}|^2}.
\label{eq:warping-radial-sum}
\end{equation}
At $|\rho_{\bm k}|=1$, the radial quotient has continuous value $N-1$, while $\mathcal C_{\bm k}^2=N$.

The first-order perturbative correction considered here changes the energies of the eigenstates, but not their wavefunctions (which are hybridized by the $\gamma_3$ term), so that the relation between amplitudes on the top and bottom surfaces remains unaltered.

\end{document}